\documentclass[11pt,a4paper]{article}
\pdfoutput=1
\usepackage{jheppub}
\usepackage{rotating}
\usepackage{array}
\usepackage{amsmath}
\usepackage{subcaption}
\usepackage[normalem]{ulem}
\usepackage{slashed}
\usepackage{booktabs}
\usepackage[utf8]{inputenc}
\usepackage[pdftex,table,dvipsnames]{xcolor}
\usepackage{units}
\usepackage{multirow}
\usepackage{url}
\usepackage{xspace}
\usepackage{color}
\DeclareUnicodeCharacter{03A5}{\ensuremath{\Upsilon}}
\preprint{TTK-19-08}

\title{ Loop-induced direct detection signatures from CP-violating scalar mediators }

\author{Fatih Ertas}
\author{and Felix Kahlhoefer}

\affiliation{ Institute for Theoretical Particle Physics and Cosmology (TTK), RWTH Aachen University, \\ D-52056 Aachen, Germany}

\emailAdd{ertas@physik.rwth-aachen.de}
\emailAdd{kahlhoefer@physik.rwth-aachen.de}

\abstract{We investigate direct detection signatures of dark matter particles interacting with quarks via a light spin-0 mediator with general CP phases. Since tree-level scattering may be strongly suppressed in the non-relativistic limit, loop contributions play an important role and can lead to observable signals in near-future experiments. We study the phenomenology of different mediator masses and CP phases with an emphasis on scenarios with maximal CP violation and Higgs portal models. Intriguingly, the sum of the rates obtained at tree- and loop-level can give a characteristic recoil spectrum not obtainable from a single type of interaction. We furthermore develop a novel method for decomposing the two-loop contribution to effective interactions between dark matter and gluons into two separate one-loop diagrams, which in our case substantially simplifies the calculation of the important top-quark contribution.}


\begin{document}

\maketitle

\flushbottom

\section{Introduction}

Experiments aiming to directly detect the interactions of dark matter (DM) particles in underground laboratories have made tremendous progress over the past decades and place some of the strongest bounds on the parameter space of many DM models~\cite{Cushman:2013zza}. Indeed, these experiments have become so sensitive that they can be relevant even for DM models where the leading order interactions are momentum- or velocity-suppressed~\cite{Xia:2018qgs}. As a result there has been a rapidly growing interest in the general effective field theory (EFT) of non-relativistic interactions between DM and nuclei~\cite{Fitzpatrick:2012ix,Anand:2013yka,Catena:2014uqa,Gresham:2014vja,Catena:2014epa,Gluscevic:2015sqa,Dent:2015zpa,Kahlhoefer:2016eds,Bishara:2016hek,Edwards:2018lsl}. In these models it becomes essential to include loop effects, which may reintroduce spin-independent interactions and thereby substantially boost the expected event rates~\cite{Haisch:2013uaa,Crivellin:2014qxa,Crivellin:2014gpa,DEramo:2016gos,Bishara:2018vix,Ghorbani:2018pjh,Yepes:2018zkk}.

Particular attention has been paid to models in which DM scattering is mediated by a pseudoscalar exchange particle~\cite{Freytsis:2010ne,Dienes:2013xya,Boehm:2014hva,Ghorbani:2014qpa,Ghorbani:2018pjh}, motivated partially by the interesting implications for collider~\cite{Ipek:2014gua,No:2015xqa,Goncalves:2016iyg,Bauer:2017ota,Bauer:2017fsw,Pani:2017qyd,Tunney:2017yfp,Banerjee:2017wxi,Haisch:2018kqx,Abe:2018bpo} and flavour~\cite{Batell:2009di,Freytsis:2009ct,Batell:2009jf,Dolan:2014ska,Berlin:2015wwa,Dobrich:2018jyi} physics. At leading order the resulting interactions are so strongly suppressed in the non-relativistic limit that they are well below the ``neutrino floor'' which indicates the ultimate reach of direct detection experiments~\cite{Billard:2013qya}. However, several recent studies have shown that loop-induced spin-independent interactions can change this picture dramatically, in particular when taking into account the interactions between the pseudoscalar mediator and the SM Higgs boson required by gauge invariance~\cite{Arcadi:2017wqi,Sanderson:2018lmj,Li:2018qip,Abe:2018emu}. In fact, ref.~\cite{Abe:2018emu} pointed out that for this particular model even two-loop processes give a relevant contribution and need to be properly included for an accurate estimate of experimental sensitivities.

In the present work we generalise these results by considering spin-0 mediators that couple to DM and Standard Model (SM) quarks with arbitrary CP phases. We furthermore treat the coupling between the mediator and SM Higgs bosons as a free parameter and thus remain agnostic about the underlying ultraviolet (UV) completion. A particular emphasis is placed on the impact of two-loop processes. We show that, at least for heavy quarks, accurate results can be obtained by first integrating out the heavy quark and then performing all further calculations in the resulting EFT. This approach substantially simplifies and speeds up the evaluation of direct detection constraints. 

We find that for general CP phases loop-induced spin-independent interactions may be strong enough to lead to detectable signals in near-future direct detection experiments, such as LZ~\cite{Akerib:2018lyp} or XENONnT~\cite{Aprile:2015uzo}. The importance of our results are illustrated for a number of relevant scenarios. We show that for DM models with maximal CP violation (as studied e.g.\ in the context of self-interacting DM~\cite{Kahlhoefer:2017umn}) loop effects can be comparable to the leading-order contribution and change the shape of the recoil spectrum in important ways. Large effects are also found in the CP-violating Higgs portal model, which has been the subject of several recent studies~\cite{Beniwal:2015sdl,Athron:2018hpc,Abe:2019wku}. In both cases loop-induced interactions enable direct detection experiments to probe parameter regions that would otherwise be out of reach.

The paper is structured as follows. In section~\ref{sec:loops} we briefly introduce the general model with free CP phases and then present our central results on how to perform the mapping onto the low-energy EFT relevant for DM direct detection. We discuss in detail the importance of two-loop processes and the matching onto non-relativistic effective operators. Specific applications of the general results are presented in section~\ref{sec:implications}, where we also calculate the sensitivity of present and future direct detection experiments. We summarise our findings and conclude in section~\ref{sec:conclusions}. Detailed results from our one-loop and two-loop calculations are presented in the appendices~\ref{app:one-loop} and~\ref{app:two-loop}, respectively. Finally, appendix~\ref{app:rel-nuc-wilson} provides details on nuclear form factors.

\section{Loop effects in direct detection}
\label{sec:loops}

We investigate a simplified model of a Dirac fermion DM particle $\chi$ interacting with SM fermions $f$ through a general spin-0 mediator $a$ with mass $m_a$ greater than the bottom-quark mass $m_b$: 
\begin{align}
\mathcal{L} = g_\chi\,a\,\bar{\chi}\, ( \cos\phi_\chi + i \gamma_5 \sin\phi_\chi)\, \chi + g_\text{SM}\sum_{f} \frac{m_f}{v} a\,\bar{f}\,( \cos\phi_\text{SM} + i \gamma_5 \sin\phi_\text{SM})\,f\;, \label{eq:L}
\end{align}
where $\phi_\chi$ and $\phi_\text{SM}$ are CP phases, $v\approx 246\,\mathrm{GeV}$ is the electroweak vacuum expectation value, $m_f$ are the SM fermion masses and $g_\chi$ as well as $g_\text{SM}$ denote the couplings of $a$ to DM and SM fermions, respectively. We have further assumed Yukawa-like couplings in agreement with the hypothesis of minimal flavour violation (MFV)~\cite{DAmbrosio:2002vsn} such that flavour physics constraints on the universal coupling $g_\text{SM}$ are weakened (see section~\ref{subsec:maxCP}).\footnote{In a generic MFV scenario a slightly more general Lagrangian than eq.~(\ref{eq:L}) can be written down, as different couplings to up- and down-type quarks are allowed. For the scope of this work, however, we will focus on the case of one universal coupling.} For $\phi_\chi = \phi_\text{SM} = 0$ we recover the well-known simplified model of a scalar mediator, whereas for $\phi_\chi = \phi_\text{SM} = \pi/2$ we obtain a CP-conserving theory with a pseudoscalar mediator~\cite{Abdallah:2015ter}. In the former case constraints on the model from direct detection experiments are very strong, whereas in the latter case they are almost entirely absent~\cite{Arcadi:2017wqi,Sanderson:2018lmj,Abe:2018emu}. Here we will treat the CP phases as free parameters in order to study the impact of different phase combinations on the predictions for direct detection experiments. 

The simplified model in eq.~(\ref{eq:L}) does not respect all gauge symmetries of the SM before electroweak symmetry breaking. The interactions between $a$ and SM fermions are therefore expected not to appear in isolation but in combination with additional interactions between $a$ and the SM Higgs boson $h$. In the present work, we will not discuss how these different interactions can be linked in specific UV completions. Instead, we introduce an additional free parameter $\lambda_{ah}$ and supplement eq.~(\ref{eq:L}) by the interaction term
\begin{align}
\mathcal{L}^\text{Higgs}_\text{int} = \frac{1}{2} \lambda_{ah} v h a^2\;.
\end{align}
We will show that this interaction can play a relevant role in the phenomenology of this model. Moreover, it will be of particular importance in section~\ref{subsec:CPHiggs} where we will identify $a$ with the SM Higgs boson $h$ itself. Note that we neglect additional interaction terms involving two Higgs bosons. Although such terms are in general expected to be present, they do not give any relevant contribution to the calculation of direct detection signatures.

We finally note that for $\phi_\text{SM} \neq \pi/2$ the mediator $a$ can mix with the SM Higgs boson, giving rise to direct interactions of the SM Higgs boson with DM particles. This mixing is however required to be small given that the observed Higgs behaves SM-like in current experiments. Furthermore, the Higgs boson mass is much larger than the values of $m_a$ that we will consider, such that its contribution to direct detection is suppressed~\cite{Sanderson:2018lmj,Abe:2018emu}. We will therefore not consider Higgs mixing within this work but emphasize that it would be straightforward to include these contributions using the results presented below.

\subsection{Low-energy effective Lagrangian}
\label{sec:one-loop}

\begin{figure}[t]
\includegraphics[width=2.75cm]{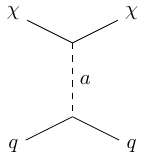}
\hspace{1.5cm}
\includegraphics[width=4.5cm]{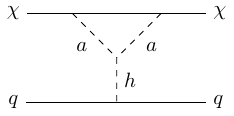}
\hspace{1.5cm}
\includegraphics[width=4.5cm]{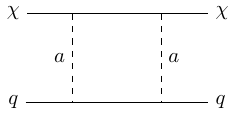}
\caption{Tree-level, Higgs-induced triangle as well as box diagram contribution to the cross section relevant for direct searches of DM. All Feynman diagrams are drawn with \texttt{TikZ-Feynman}~\cite{Ellis:2016jkw}.}\label{fig:CPDiagram1}
\end{figure}

\noindent
To calculate event rates in direct detection experiments, we need to determine the effective interactions between DM and quarks that result from the three types of diagrams illustrated in figure~\ref{fig:CPDiagram1}. For the discussion below it will be useful to distinguish between interactions that lead to spin-independent (SI) and to spin-dependent (SD) scattering in the non-relativistic limit.\footnote{Note that here and below we use the term ``spin-independent'' to refer to all types of interactions that do not depend on the nucleus spin, irrespective of whether or not they are suppressed in the non-relativistic limit. Accordingly, the term ``spin-dependent'' refers to all interactions that are not spin-independent, including momentum-suppressed interactions. Indeed, unsuppressed spin-dependent interactions are absent in the model that we consider.} Starting with the tree-level exchange of $a$ illustrated in the left panel of figure \ref{fig:CPDiagram1}, we obtain
\begin{align}
\mathcal{L}^{\text{SI}}_{\text{tree}} &= \sum_{q=\text{all}} m_q\,\mathcal{C}^\text{tree}\big(\cos(\phi_\chi) \cos(\phi_\text{SM}) \,\bar{\chi} \chi + \sin(\phi_\chi) \cos(\phi_\text{SM})\,\bar{\chi}i \gamma_5 \chi\big)\,\bar{q} q\;,\\
\mathcal{L}^{\text{SD}}_{\text{tree}} &= \sum_{q=\text{all}} m_q\,\mathcal{C}^\text{tree}\big(\cos(\phi_\chi) \sin(\phi_\text{SM}) \,\bar{\chi} \chi + \sin(\phi_\chi) \sin(\phi_\text{SM})\,\bar{\chi}i \gamma_5 \chi\big)\,\bar{q}i \gamma_5 q\;,
\end{align}
where the sum runs over all quark species. Here we have defined the tree-level coefficient
\begin{align}
\mathcal{C}^\text{tree} = \frac{g_\chi\, g_\text{SM}}{v\, m_a^2}\;,
\end{align}
and have kept the dependence on the two CP phases explicit.

Next we consider the Higgs-mediated exchange shown in the middle panel of figure \ref{fig:CPDiagram1}, which maps onto the purely spin-independent interaction
\begin{align}
\mathcal{L}^\text{SI}_{\text{triangle}} =\sum_{q=\text{all}} \frac{m_q \lambda_{ah}}{m_h^2}\left( \mathcal{C}_S^{\text{triangle}} \,\bar{\chi} \chi\,\bar{q} q +\mathcal{C}_{PS}^{\text{triangle}} \,\bar{\chi} i \gamma_5 \chi\, \bar{q} q\right) \;,
\end{align}
where the sum again runs over all quarks. We have further introduced the triangle coefficients
\begin{align}\label{eq:trianglecoeff1}
\mathcal{C}_S^{\text{triangle}} &= \frac{g_\chi^2}{(4\pi)^2}\, m_\chi \left[ (1 + \cos(2\phi_\chi))\,C_0(m_\chi^2,\, m_a^2,\,m_\chi^2) + C_2(m_\chi^2,\, m_a^2,\,m_\chi^2)\right]\;,\\
\label{eq:trianglecoeff2}
\mathcal{C}_{PS}^{\text{triangle}} &= \frac{g_\chi^2}{(4\pi)^2}\, m_\chi \sin(2\phi_\chi)\,C_0(m_\chi^2,\, m_a^2,\,m_\chi^2)\;,
\end{align}
in terms of the loop functions $C_0(m_\chi^2,\, m_a^2,\,m_\chi^2)$ and $C_2(m_\chi^2,\, m_a^2,\,m_\chi^2)$, which are given in appendix~\ref{app:one-loopfcts}. 

Finally, we have to take into account the box diagram in the right panel of figure~\ref{fig:CPDiagram1}. We expand the amplitude in terms of the quark momentum, which is the smallest scale in the diagram~\cite{Abe:2018emu}, and obtain
\begin{align}
\begin{split}
\label{effbox1}
\mathcal{L}^{\text{SI}}_{\text{box}} =& \sum_{q=u,d,s} \left(m_q\,\mathcal{C}^\text{box}_{1,q}\,\bar{\chi} \chi \,\bar{q} q + m_q\,\mathcal{C}^\text{box}_{2,q} \,\bar{\chi} i \gamma_5 \chi\,\bar{q} q\right)\\
 &+\sum_{q=u,d,s,c,b}\Big(\mathcal{C}^\text{box}_{5,q} \,\bar{\chi} i\partial^\mu \gamma^\nu \chi \, \mathcal{O}^q_{\mu\nu}
+ \mathcal{C}^\text{box}_{6,q} \,\bar{\chi} i\partial^\mu i \partial^\nu \chi \, \mathcal{O}^q_{\mu\nu}+\mathcal{C}^\text{box}_{7,q} \,\bar{\chi} i \gamma_5 i\partial^\mu i \partial^\nu \chi \, \mathcal{O}^q_{\mu\nu}\Big)\;,
\end{split}\raisetag{3.3\normalbaselineskip}\\
\label{effbox2}
\mathcal{L}^{\text{SD}}_{\text{box}} =&  \sum_{q=u,d,s} \left(m_q\,\mathcal{C}^\text{box}_{3,q}\,\bar{\chi} \chi \,\bar{q}i \gamma_5 q + m_q\,\mathcal{C}^\text{box}_{4,q}\,\bar{\chi} i \gamma_5 \chi \,\bar{q} i \gamma_5 q\right)\;.
\end{align}
Computational details and the expressions of the different box diagram coefficients $\mathcal{C}^\text{box}_{i,q}$ are given in appendix~\ref{app:one-loopwilson}. Note that all of these coefficients share a common factor of $g^2_\chi\, g^2_\text{SM}\,m_q^2/v^2$, which also constitutes the only quark dependence. In eq.~(\ref{effbox1}) we have also introduced the twist-2 quark operator 
\begin{align}
\mathcal{O}^q_{\mu\nu} = \bar{q}\,\left(\frac{i \partial^\mu \gamma^\nu + i \partial^\nu \gamma^\mu}{2} - \frac{1}{4} g^{\mu\nu} i \slashed{\partial}\right)\,q\;.
\end{align}
Since the corresponding form factors are evaluated at the scale of the $Z$ boson mass $m_Z$, we include the charm and bottom quark in the corresponding sums in eq.~(\ref{effbox1})~\cite{Hisano:2010ct,Hisano:2015bma}. However, none of the heavy quarks have been included in the remaining terms of eqs.~(\ref{effbox1}) and~(\ref{effbox2}), because they require a different treatment, which will be discussed next.

\subsection{Effective description of two-loop processes}\label{subsec:eff2Loop}

\begin{figure}[t]
\centering
\begin{subfigure}{0.24\textwidth}
\includegraphics[width=\textwidth]{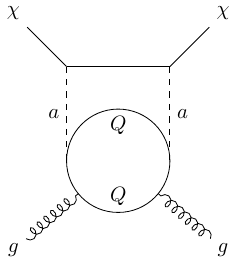}
\end{subfigure}
\qquad
\begin{subfigure}{0.24\textwidth}
\includegraphics[width=\textwidth]{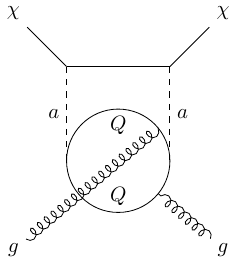}
\end{subfigure}
\caption{Two-loop processes for the evaluation of the heavy quark ($Q = c,b,t$) contributions to effective DM-gluon interactions.}\label{fig:CPDiagram2}
\end{figure}

As the charm, bottom and top quark are heavier than the energy scale relevant for DM direct detection experiments, they should be integrated out of the theory aiming to describe interactions at the level of nuclei. For the tree-level and Higgs-induced triangle diagram this can be done simply by replacing the heavy quarks by the corresponding effective gluon interaction obtained from triangular heavy-quark loops~\cite{Shifman:1978zn}
\begin{align}
\label{eq:QQGGShifman}
m_Q \bar{Q} Q &\rightarrow -\frac{\alpha_s}{12\pi} G^{a}_{\mu\nu} G^{a\mu\nu}\;,\\
m_Q \bar{Q} i \gamma_5 Q &\rightarrow \frac{\alpha_s}{8\pi} G^{a}_{\mu\nu} \widetilde{G}^{a\mu\nu}\;,
\end{align}
where $G^{a\mu\nu}$ is the gluon field strength tensor and ${\widetilde{G}}^{a\mu\nu} = \frac{1}{2} \epsilon^{\mu\nu\alpha\beta}G^a_{\alpha\beta}$ with the convention $\epsilon^{0123} = 1$. This procedure is justified for these two diagrams since the two steps of integrating out the mediator $a$ and integrating out the heavy quarks factorise. 

The situation is however very different for the box diagram in the right panel of figure~\ref{fig:CPDiagram1}. In this case one cannot make a simple factorization argument to integrate out heavy quarks. This is visualised in figure~\ref{fig:CPDiagram2}, which shows the two-loop diagrams that need to be computed to obtain the effective interactions between DM and gluons. Any attempt to simplify this calculation by first integrating out the mediator $a$ and then using eq.~(\ref{eq:QQGGShifman}) would neglect the contribution from the diagram on the right. For $m_Q \ll m_a,\,m_\chi$ the two-loop computation hence cannot be simplified in this way without introducing potentially large errors~\cite{Abe:2018emu}. In the opposite case of $m_Q \gg m_a,\,m_\chi$ it was argued in ref.~\cite{Abe:2018emu} that a simplification is not possible because one cannot expand the box diagram amplitude in terms of the external quark momentum, which is no longer the smallest scale in the diagram. It was in particular stressed that for the top quark a full two-loop computation is mandatory.

\begin{figure}[t]
\begin{subfigure}{0.24\textwidth}
\includegraphics[width=\textwidth]{Full2Loop.pdf}
\end{subfigure}
{+}
\begin{subfigure}{0.24\textwidth}
\includegraphics[width=\textwidth]{Full2LoopCrossed.pdf}
\end{subfigure}
{$\rightarrow$}
\begin{subfigure}{0.24\textwidth}
\includegraphics[width=\textwidth]{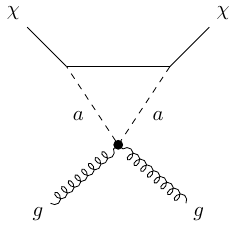}
\end{subfigure}
{$\rightarrow$}
\begin{subfigure}{0.16\textwidth}
\includegraphics[width=\textwidth]{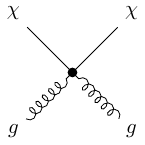}
\end{subfigure}
\caption{Illustration of the decomposition of the two-loop process for \mbox{$m_Q \gg m_a,\,m_\chi$}. After first integrating out the heavy quark $Q$ (first arrow) one can then match the resulting one-loop diagram onto effective DM-gluon interactions (second arrow). The black dot represents an effective interaction corresponding to a higher-dimensional operator.}\label{fig:CPDiagram4}
\end{figure}

As we are now going to demonstrate, however, for $m_Q \gg m_a,\,m_\chi$ it is in fact possible to decompose the underlying two-loop process into two separate one-loop diagrams by integrating out the \mbox{heavy quark $Q$} first and the mediator $a$ afterwards. This approach, in which no diagrams are neglected, is illustrated in figure~\ref{fig:CPDiagram4}. Provided the mediator is light compared to the heavy quark, it is thus possible to simplify the calculations significantly.

In the following we will be mostly interested in the case $m_a \ll m_t$, such that the approach outlined above can be applied to the top quark. Therefore, we first consider the loop involving the top quark separately and integrate it out by performing a $1/m_t$ expansion of the (in total six) corresponding amplitudes. We employ \mbox{\texttt{Package-X}~\cite{Patel:2015tea}} for the evaluation and expansion of the loop computations. This then yields the following leading order effective Lagrangian coupling $a$ to gluons
\begin{align}
\label{effaaGMatch}
\mathcal{L}_{\text{eff}}^{\text{aaG}} &= \frac{1}{2}\,d^\text{\,eff}_G\,a a\,\frac{\alpha_s}{12\pi}\,G^{a}_{\mu\nu} G^{a\mu\nu} +\frac{1}{2}\,d^\text{\,eff}_{\widetilde{G}}\,a a\,\frac{\alpha_s}{8\pi}G^{a}_{\mu\nu} \widetilde{G}^{a\mu\nu}\;.
\end{align}
Here we have included a symmetry factor of 1/2 and defined\footnote{Note that $d^\text{\,eff}_{G}$ vanishes for certain values of $\phi_\text{SM}$ such that one would need to include higher orders. However, these specific cases are not of interest in the present work. While $d^\text{\,eff}_{\widetilde{G}}$ also vanishes for specific values of $\phi_\text{SM}$, the same is true for the full expression $d^\text{\,full}_{\widetilde{G}}$, see eq.~(\ref{app:dGdualFull}) in appendix~\ref{app:two-loopcomp}, i.e.\ this is not a result of the heavy quark expansion.}
\begin{align}
d^\text{\,eff}_G = - \frac{g_\text{SM}^2}{v^2} & \cos(2 \phi_\text{SM})\;,\qquad d^ \text{\,eff}_{\widetilde{G}} = \frac{g_\text{SM}^2}{v^2} \sin(2 \phi_\text{SM})\;,
\end{align}
which are both independent of the top-quark mass. Now performing the second step visualised in figure \ref{fig:CPDiagram4}, we obtain for the effective two-loop approach
\begin{align}
\mathcal{L}^\text{SI}_{\text{2-Loop}} &= \left(\mathcal{C}^\text{eff}_{G,S}\, \bar{\chi} \chi+\mathcal{C}^\text{eff}_{G,PS}\, \bar{\chi} i \gamma_5 \chi\,\right)\frac{-\alpha_s}{12\pi}\,G^{a}_{\mu\nu} G^{a\mu\nu}\;,\\
\mathcal{L}^\text{SD}_{\text{2-Loop}} &= \left(\mathcal{C}^\text{eff}_{\widetilde{G},S}\, \bar{\chi} \chi+\mathcal{C}^\text{eff}_{\widetilde{G},PS}\, \bar{\chi} i \gamma_5 \chi\, \right)\frac{\alpha_s}{8\pi}\,G^{a}_{\mu\nu} {\widetilde{G}}^{a\mu\nu}\;,
\end{align}
where the effective two-loop coefficients read
\begin{align}
\mathcal{C}^\text{eff}_{G,S} &= d^\text{\,eff}_G\,\mathcal{C}_S^{\text{triangle}}\;,\qquad &\mathcal{C}^\text{eff}_{G,PS} &= d^\text{\,eff}_G\, \mathcal{C}_{PS}^{\text{triangle}}\;,\\
\mathcal{C}^\text{eff}_{G,S}&= -d^\text{\,eff}_{\widetilde{G}}\, \mathcal{C}_S^{\text{triangle}}\;,\qquad &\mathcal{C}^\text{eff}_{\widetilde{G},PS} &= -d^\text{\,eff}_{\widetilde{G}}\, \mathcal{C}_{PS}^{\text{triangle}}\;.
\end{align}

An analogous calculation for the bottom and charm quark only gives a useful approximation if $m_a \ll m_c,\,m_b$. For heavier mediator masses it is in general unavoidable to perform the full two-loop calculation to accurately estimate the corresponding contributions (see appendix~\ref{app:two-loopcomp} for more details). However, for the specific coupling structure that we are interested in, bottom and charm quark are found to give only a small contribution.\footnote{This conclusion could change for example in models with extended Higgs sectors, where couplings to down-type quarks may receive a substantial enhancement.} It is hence possible to obtain a very good approximate result of the total heavy quark contribution to the effective DM-gluon interactions by including only the top-quark contribution using our effective approach.

\begin{figure}[t]
\centering
\includegraphics[width=6.8cm]{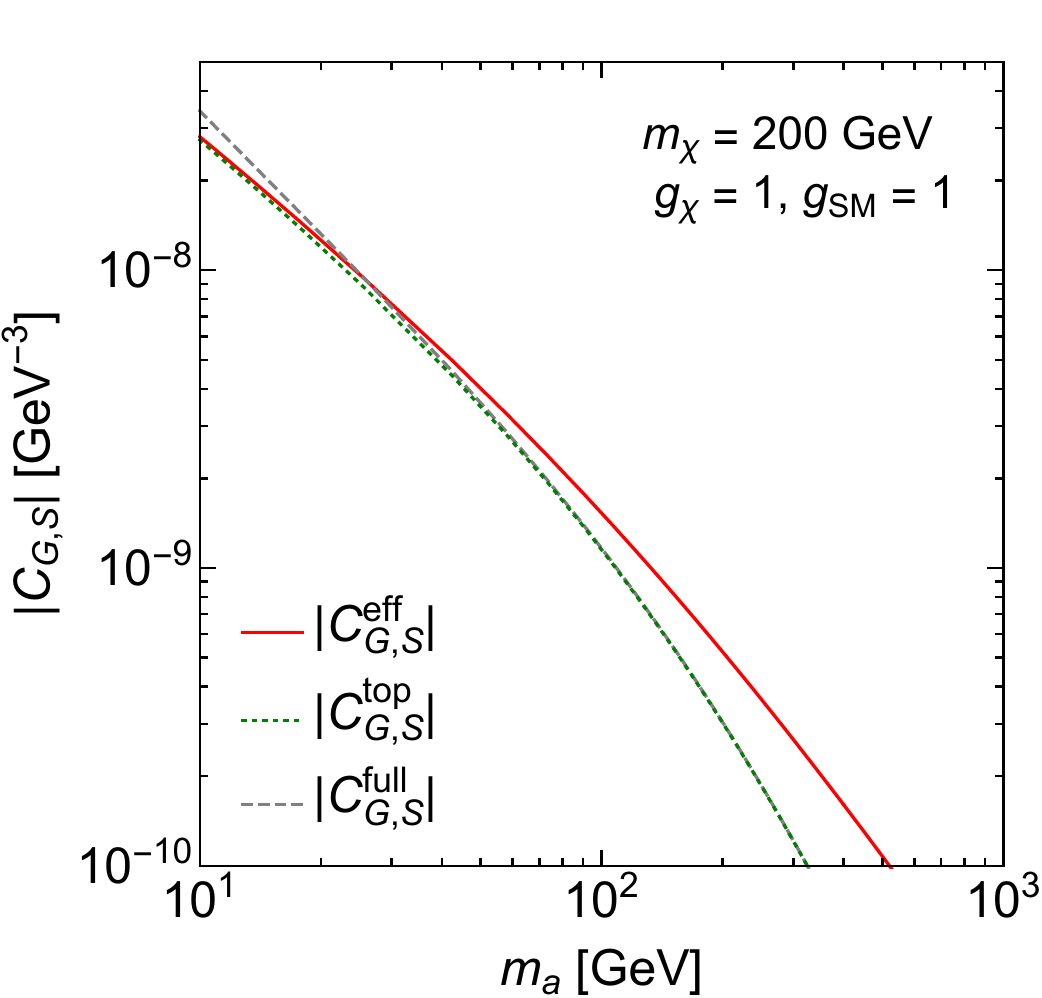}
\hspace{1cm}
\includegraphics[width=6.8cm]{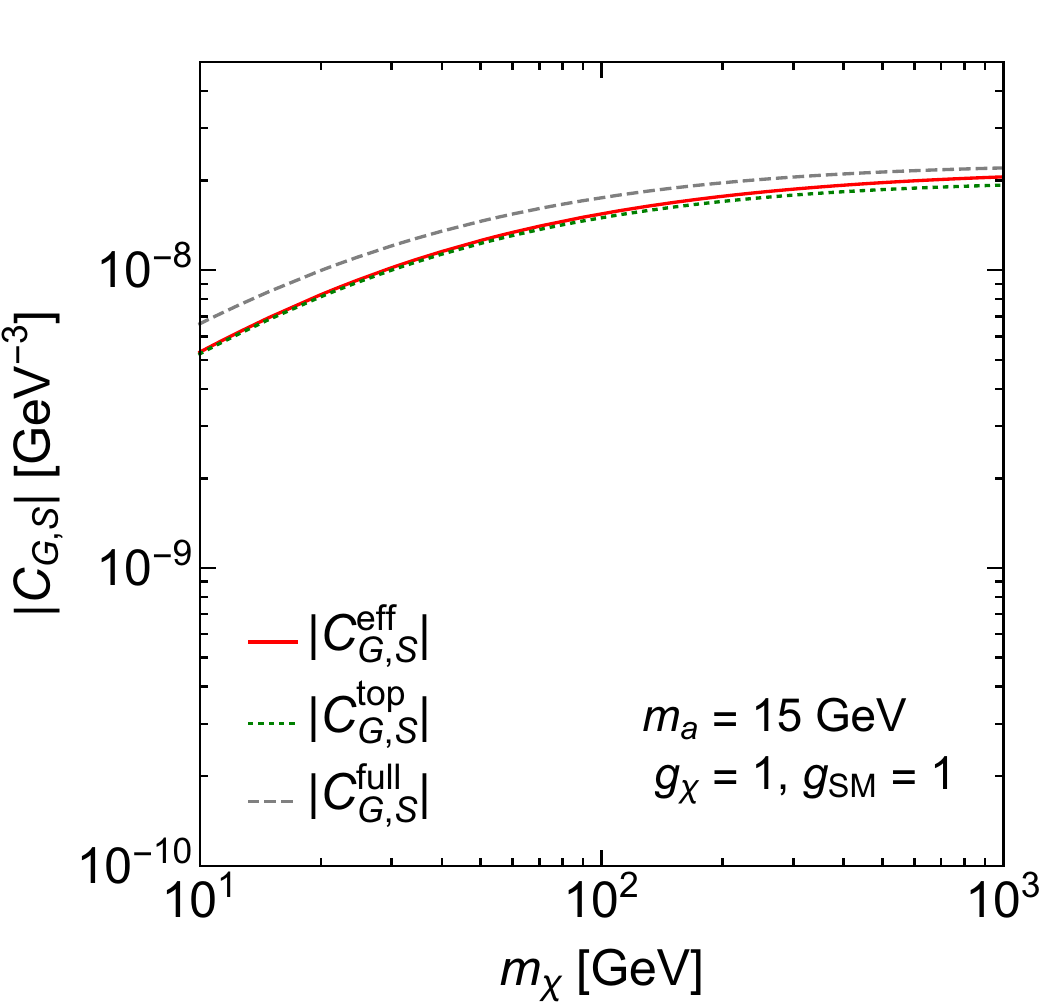}
\caption{Comparison of $|\mathcal{C}_{G,S}|$ in the effective approach for the top quark (red), the two-loop contribution of the top quark (dotted green) and the two-loop result including all heavy quarks (dashed grey) as a function of $m_a$ (left panel) and $m_\chi$ (right panel). For both plots we set $\phi_\chi = 0$ and $\phi_\text{SM} = \pi/2$.}\label{fig:CPComparisonCoeff}
\end{figure}

This is illustrated in figure~\ref{fig:CPComparisonCoeff}, where we plot the absolute value of the coefficient $\mathcal{C}_{G,S}$ as a function of the mediator mass (left panel) and of the DM mass $m_\chi$ (right panel). The effective approach for the top quark (indicated by the solid red line) and the corresponding two-loop calculation (dotted green) show very good agreement for $m_a \ll m_t$ across the whole range of DM masses. Including also bottom and charm quark in the two-loop calculation has only slight influences for small values of $m_a$ (dashed grey). Similar results can be obtained for the other coefficients.\footnote{For specific parameter points cancellations might occur within the coefficients $\mathcal{C}_{G}$ and $\mathcal{C}_{\widetilde{G}}$ like in $d_G^\text{\,eff}$ for $\phi_\text{SM} \approx \pi/4$. In this parameter region the two-loop result and the effective approach differ. However, this discrepancy does not affect any of the scenarios studied in detail below.} We conclude that it is possible to simplify the full two-loop calculation in the case of $m_Q \gg m_a,\,m_\chi$, which makes it possible to circumvent the full two-loop calculation entirely if the top quark is expected to give the dominant contribution. We will therefore use the effective approach for the remainder of this work.

\subsection{Matching onto effective operators}\label{subsec:Matching}
In this section we match the effective interactions of DM with quarks and gluons onto non-relativistic DM-nucleon interactions in order to obtain predictions for direct detection experiments. The first step is to perform the matching of quark and gluon fields onto nucleon fields, which yields the following effective Lagrangian:
\begin{align}
\mathcal{L}_{\chi N}^{\text{eff}} &= \left(\mathcal{C}^{\text{SI}}_{\text{eff},N}\,\bar{\chi}\chi  + \mathcal{C}^{\text{SI,CPV}}_{\text{eff},N}\, \bar{\chi}i\gamma_5\chi\right) \bar{N} N 
+\left(\mathcal{C}^{\text{SD,CPV}}_{\text{eff},N}\, \bar{\chi}\chi  +\mathcal{C}^{\text{SD}}_{\text{eff},N}\,\bar{\chi}i\gamma_5\chi\right) \bar{N} i\gamma_5N\;, \label{eq:LeffchiN}
\end{align}
where $N = p,n$ is a nucleon field and `CPV' indicates terms that only arise when CP is violated. The coefficients $C_\text{eff}$ depend on the various coefficients we derived in the previous two sections as well as on the nuclear form factors that parametrise the quark and gluon contents of a nucleon. Note that in general the nuclear form factors and hence the effective coefficients are different for protons and neutrons: $C_{\text{eff},p} \neq C_{\text{eff},n}$.

For the SI coefficients, we find
\begin{align}
\notag
\mathcal{C}^{\text{SI}}_{\text{eff},N}  &=\sum_{q=u,d,s} m_N f^N_q\left( \cos(\phi_\chi) \cos(\phi_\text{SM})\,\mathcal{C}^\text{tree}  + \frac{\lambda_{ah}}{m_h^2} \,\mathcal{C}^\text{triangle}_{S}+\mathcal{C}^\text{box}_{1,q} \right)\\
&\hspace{0.5cm}+ 3\cdot \frac{2}{27} m_N f^N_G\left( \cos(\phi_\chi) \cos(\phi_\text{SM})\,\mathcal{C}^\text{tree} + \frac{ \lambda_{ah}}{m_h^2} \,\mathcal{C}^\text{triangle}_{S} \right)\\
\notag
&\hspace{0.5cm}+ \sum_{q=u,d,s,c,b}  \frac{3}{4} m_N m_\chi \,\Big(q^{N}(2) + \bar{q}^{N}(2)\Big) \Big(\mathcal{C}^\text{box}_{5,q} + m_\chi\,\mathcal{C}^\text{box}_{6,q}\Big) + \frac{2}{27} m_N f^N_G\,\mathcal{C}^\text{eff}_{G,S}\;,
\end{align}
as well as
\begin{align}
\notag
\mathcal{C}^{\text{SI,CPV}}_{\text{eff},N} &= \sum_{q=u,d,s} m_N f^N_q\left( \sin(\phi_\chi) \cos(\phi_\text{SM})\,\mathcal{C}^\text{tree}  +  \frac{\lambda_{ah}}{m_h^2}\,\mathcal{C}^\text{triangle}_{PS}+\mathcal{C}^\text{box}_{2,q}  \right)\\
&\hspace{0.5cm}+ 3 \cdot \frac{2}{27} m_N f^N_G \left( \sin(\phi_\chi) \cos(\phi_\text{SM})\,\mathcal{C}^\text{tree} + \frac{\lambda_{ah}}{m_h^2}\,\mathcal{C}^\text{triangle}_{PS} \right)\\
\notag
&\hspace{0.5cm} + \sum_{q=u,d,s,c,b}  \frac{3}{4} m_N m^2_\chi \,\Big(q^{N}(2) + \bar{q}^{N}(2)\Big)\,\mathcal{C}^\text{box}_{7,q} + \frac{2}{27} m_N f^N_G \,\mathcal{C}^\text{eff}_{G,PS}\;,
\end{align}
where the nuclear form factors $f^N_{q,G}$, $q^N(2)$ and $\bar{q}^N(2)$ are defined in appendix~\ref{app:rel-nuc-wilson}. 

Likewise, we obtain for the SD coefficients
\begin{align}
\begin{split}
 \mathcal{C}^{\text{SD,CPV}}_{\text{eff},N}&= \sum_{q=u,d,s} F_P^{q/N} \,\Big(\cos(\phi_\chi) \sin(\phi_\text{SM}) \,\mathcal{C}^\text{tree} + \mathcal{C}^\text{box}_{3,q}\Big)\\
&\hspace{0.5cm}+\, F^N_{\widetilde{G}}\, \Big(3 \cos(\phi_\chi) \sin(\phi_\text{SM}) \,\mathcal{C}^\text{tree} + \,\mathcal{C}^\text{eff}_{\widetilde{G},S}  \Big)\;,
\end{split}
\raisetag{3.4\normalbaselineskip}
\end{align}
and
\begin{align}
\begin{split}
\mathcal{C}^{\text{SD}}_{\text{eff},N} &= \sum_{q=u,d,s} F_P^{q/N} \,\Big(\sin(\phi_\chi) \sin(\phi_\text{SM}) \,\mathcal{C}^\text{tree} + \mathcal{C}^\text{box}_{4,q}\Big)\\
&\hspace{0.5cm}+\, F^N_{\widetilde{G}}\, \Big(3 \sin(\phi_\chi) \sin(\phi_\text{SM}) \,\mathcal{C}^\text{tree} + \,\mathcal{C}^\text{eff}_{\widetilde{G},PS}  \Big)\;.
\end{split}
\raisetag{3.4\normalbaselineskip}
\end{align}
The form factors $F_P^{q/N}$ and $F^N_{\widetilde{G}}$ are given in appendix~\ref{app:rel-nuc-wilson}. Because of non-negligible contributions from the $\pi$ and $\eta$ pole, these form factors depend on the momentum exchange $q^\mu$ between DM and nucleons.

In the non-relativistic limit the effective Lagrangian from eq.~(\ref{eq:LeffchiN}) can be matched onto a basis of effective operators:
\begin{equation}
\mathcal{L}_{\chi N}^{\text{eff}} \to \sum_i c_i^N \mathcal{O}^N_i \; ,
\end{equation}
where the operators $\mathcal{O}^N_i$ depend only on the spins $\vec{S}_\chi$ and $\vec{S}_N$ of DM and the nucleon, respectively, as well as on the momentum transfer $\vec{q}$ and the DM-nucleon relative velocity $\vec{v}$~\cite{Fitzpatrick:2012ix,Anand:2013yka,Fan:2010gt}. For the model that we consider, only four different operators are generated, namely
\begin{align}
\begin{split}
 \mathcal{O}^N_1 & = 1 \, ,\\ \mathcal{O}^N_6 & = (\vec{S}_\chi \cdot \frac{\vec{q}}{m_N}) (\vec{S}_N \cdot \frac{\vec{q}}{m_N}) \, ,  \\ \mathcal{O}^N_{10} & = i (\vec{S}_N \cdot \frac{\vec{q}}{m_N}) \, ,  \\ \mathcal{O}^N_{11} & = i(\vec{S}_\chi \cdot \frac{\vec{q}}{m_N}) \; .
\end{split}
\end{align}
The corresponding coefficients can be directly read off from $\mathcal{L}_{\chi N}^\text{eff}$~\cite{Anand:2013yka}:
\begin{equation}
\label{eq:NRcoeff}
 c_1^N = C_{\text{eff},N}^\text{SI} \, , \qquad  c_6^N = \frac{m_N}{m_\chi} C_{\text{eff},N}^\text{SD} \, , \qquad c_{10}^N = C_{\text{eff},N}^\text{SD,CPV} \, , \qquad c_{11}^N = - \frac{m_N}{m_\chi} C_{\text{eff},N}^\text{SI,CPV} \; .
\end{equation}
Note that like the form factors $F_P^{q/N}$ and $F^N_{\widetilde{G}}$ the coefficients $c_6^N$ and $c_{10}^N$ also depend on the momentum transfer. This final step completes the derivation of the effective interactions relevant for DM direct detection from the general Lagrangian of a spin-0 mediator given in eq.~(\ref{eq:L}).

\section{Phenomenological implication}
\label{sec:implications}

In this section we use the results from above to predict the differential event rates in past and future direct detection experiments and to calculate the resulting exclusion limits and expected sensitivities. In models that predict dominantly spin-independent scattering, this can be done by simply calculating the corresponding scattering cross section
\begin{equation}
 \sigma^\text{SI}_N = \frac{\mu_N^2 \, |c_1^N|^2}{\pi} \; ,
\end{equation}
where $\mu_N = m_\chi m_N / (m_\chi + m_N)$ is the DM-nucleon reduced mass. For $c_1^p \approx c_1^n$ the differential event rate with respect to recoil energy $E_\mathrm{R}$ is then simply given by 
\begin{align}
\label{dRdER}
\frac{\mathrm{d}R}{\mathrm{d}E_\mathrm{R}} = \frac{\rho_0 \, \sigma^\text{SI}_p \, A^2 \, F^2(E_{\text{R}})}{2 \, \mu_p^2 \, m_\chi} g(v_\text{min}) \; ,
\end{align}
where $\rho_0$ is the local DM density, $A$ is the mass number of the target nucleus and $F^2(E_\text{R})$ denotes the nuclear form factor. The factor $g(v_\text{min}) = \int_{v_\text{min}} f(v)/v \, \mathrm{d}v$ denotes the velocity integral as a function of the minimum velocity $v_\text{min}(E_\text{R}) = \sqrt{m_A E_\text{R} / (2 \, \mu^2)}$ with $m_A$ being the mass of the target nucleus and $\mu$ being the corresponding reduced mass. Direct detection experiments typically assume this particular form of the differential cross section in order to produce exclusion limits and quote expected sensitivities in terms of $\sigma^\text{SI}_p$ as a function of $m_\chi$.

In the presence of additional interactions, however, the calculation of the differential event rate becomes much more involved. We do not review the corresponding formalism here and instead refer to refs.~\cite{Fitzpatrick:2012ix,Anand:2013yka,Kahlhoefer:2016eds}. Crucially, for momentum-dependent interactions it is no longer possible to capture the model prediction in terms of a single cross section at fixed momentum transfer which can then be compared to published exclusion limits. To evaluate experimental sensitivity it thus becomes necessary to reproduce experimental analyses for the appropriate recoil spectra and include information on detection efficiencies and background levels in order to obtain approximate likelihood functions.

This process has been automated for the most general set of non-relativistic effective operators in the public code \texttt{DDCalc\_v2.1}~\cite{Workgroup:2017lvb,Athron:2018hpc}, which includes an extensive database of existing and planned direct detection experiments. Furthermore, \texttt{DDCalc} contains an automated interface with \texttt{DirectDM}~\cite{Bishara:2017nnn}, which we use for the matching of the spin-dependent coefficients in eq.~(\ref{eq:NRcoeff}) and the evaluation of the corresponding nuclear form factors. We can therefore simply pass the coefficients $C_{\text{eff},N}$ calculated for our model to \texttt{DDCalc} and obtain the likelihoods for existing direct detection experiments and the predicted number of events in future experiments. In the following, we will indicate the regions of parameter space that are excluded by the most recent XENON1T results~\cite{Aprile:2018dbl} and the regions that predict at least 5 events in the next-generation LZ experiment~\cite{Akerib:2018lyp}.\footnote{This number of events corresponds approximately to the median expected sensitivity using a cut-and-count analysis with a background expectation of 6.49 events. A better sensitivity may be achieved by exploiting differences in the differential distributions between signal and background.} Similar exclusion limits are obtained from the Panda-X~\cite{Cui:2017nnn,Xia:2018qgs} and LUX~\cite{Akerib:2016vxi,Akerib:2017kat} experiments, while comparable sensitivities are expected for the XENONnT experiment~\cite{Aprile:2015uzo}. 

\subsection{General CP phases}\label{subsec:genCP}

\begin{figure}[t]
  \centering
  \includegraphics[width=0.49\textwidth]{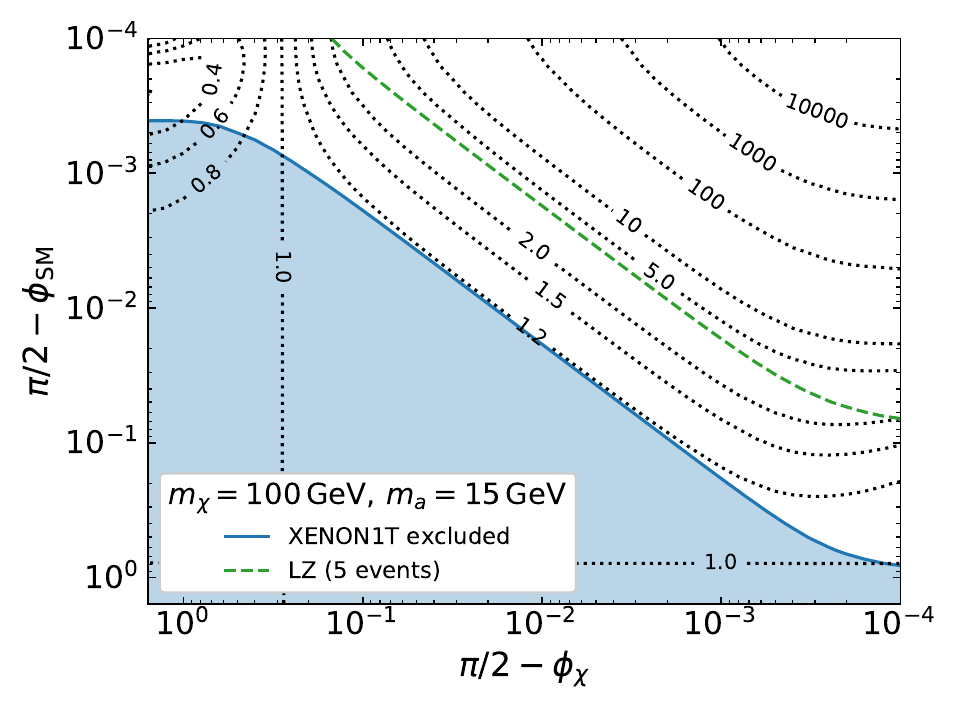}
  \includegraphics[width=0.49\textwidth]{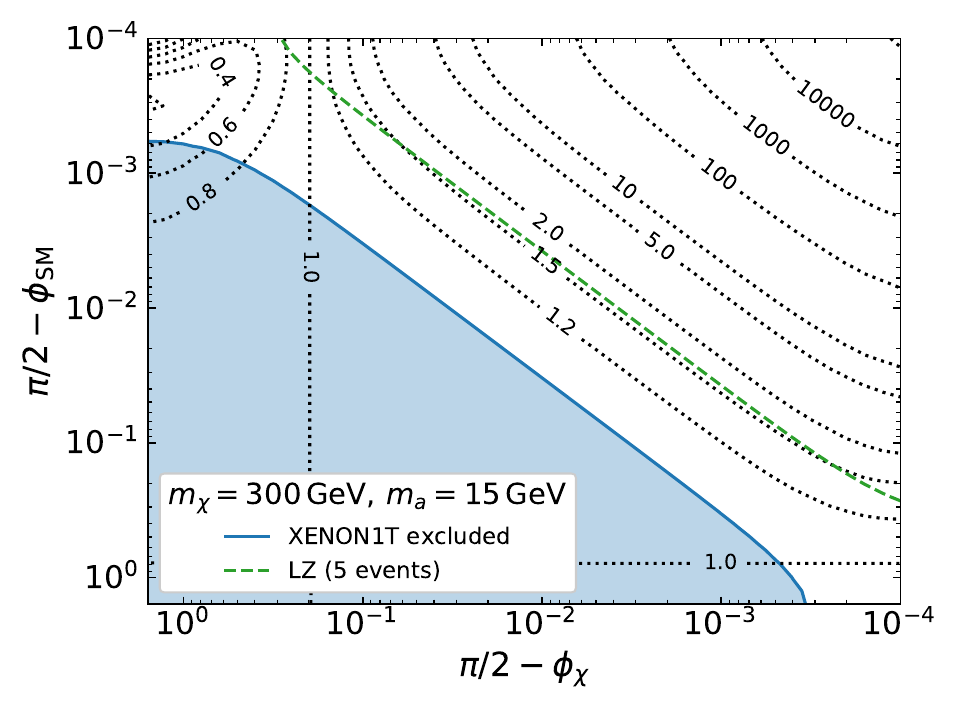}
  \caption{
  Direct detection constraints as a function of the CP-violating phases $\phi_\chi$ and $\phi_\text{SM}$. For both panels we have fixed $g_\chi = g_\text{SM} = 1$ and $\lambda_{ah} = 0$. The dotted lines indicate the ratio of the predicted number of events in LZ when including both tree-level and loop-level diagrams and when only including tree-level diagrams. This ratio can be smaller than unity due to destructive interference.}
  \label{fig:general}
\end{figure}

We first visualize the general aspects of our model by considering in figure~\ref{fig:general} the most general case, in which $\phi_\chi$ and $\phi_\text{SM}$ can take arbitrary values between 0 (corresponding to purely scalar couplings) and $\pi/2$ (corresponding to purely pseudoscalar couplings). For the purpose of this figure we have fixed $g_\text{SM} = g_\chi = 1$ and $\lambda_{ah} = 0$ and consider two different combinations of $m_\chi$ and $m_a$ in the two panels. Note that we assume that the correct relic density is reproduced at each point in both plots without invoking a specific mechanism. The blue shading indicates the parameter region excluded by XENON1T, while the dashed green line provides an estimate for the reach of LZ. The black dotted lines indicate the ratio of the total number of predicted events in LZ to the number of events predicted at tree-level. 

For $\phi_\chi$, $\phi_\text{SM} \ll \pi/2$ the tree-level exchange of $a$ dominates the spin-independent coefficient $\mathcal{C}^{\text{SI}}_{\text{eff},N}$ from eq.~(\ref{eq:LeffchiN}) and therefore also the whole scattering process. In such a scenario current direct detection bounds rule out a large part of the parameter space and constrain $g_\text{SM}$ to be very small~\cite{Kaplinghat:2013yxa,Kahlhoefer:2017umn}. As the two phases approach $\pi/2$, tree-level scattering becomes more and more suppressed, leading to a reduced sensitivity of direct detection experiments and a greater importance of loop effects. 

For $\phi_\chi = 0$ and \mbox{$\phi_\text{SM} = \pi/2$}, i.e.\ the top-left corner of figure~\ref{fig:general}, CP violation is maximal. In this case the tree-level contribution maps onto the non-relativistic operator $\mathcal{O}^N_{10}$, which is suppressed in the non-relativistic limit and furthermore depends on the spin of the nucleus. Existing direct detection constraints can thus be evaded even with $\mathcal{O}(1)$ couplings~\cite{Dienes:2013xya}. However, spin-independent contributions arise at loop-level and can dominate the event rate and yield potentially observable signals. The importance of loop-effects can also be seen for $\phi_\chi \approx 0$ and $10^{-4} \lesssim \pi/2 - \phi_\text{SM} \lesssim 10^{-3}$, where the total event rate is \emph{smaller} than the one predicted at tree-level due to the destructive interference between spin-independent interactions present at tree-level and those induced at loop-level. We discuss the case of maximal CP violation in more detail in section~\ref{subsec:maxCP}. 

For the opposite scenario of $\phi_\chi = \pi/2$ and $\phi_\text{SM} = 0$, i.e.\ the bottom-right corner of figure~\ref{fig:general}, the tree-level contribution to spin-independent scattering maps onto the non-relativistic operator $\mathcal{O}^N_{11}$, which depends on the DM spin and the momentum transfer. While the scattering cross section does receive a coherent enhancement in this case, it is suppressed by an additional factor of $m_N^2 / m_\chi^2$. We will therefore study the influence of purely spin-independent contributions emerging at loop-level in the context of the CP-violating Higgs-portal model in section~\ref{subsec:CPHiggs}. 

Finally, in the top-right corner of figure~\ref{fig:general}, corresponding to almost purely pseudoscalar interactions, the loop-induced event rate dominates over the tree-level prediction by many orders of magnitude. However, as observed previously~\cite{Abe:2018emu}, the sensitivity of direct detection experiments is strongly suppressed in this limit, so that the case of pure pseudoscalar interactions is out of reach for current direct detection experiments.
A crucial conclusion from figure~\ref{fig:general} is that loop effects become increasingly important as experimental sensitivity improves. For the couplings and masses considered, XENON1T is only sensitive to those regions in parameter space where loop-induced interactions give a sub-leading contribution. LZ on the other hand will be sensitive to interactions that are more strongly suppressed at tree-level, giving greater importance to an accurate calculation of loop-level contributions.

\subsection{Maximal CP violation}\label{subsec:maxCP}

Let us take a closer look at the case $\phi_\chi = 0$ and $\phi_\text{SM} = \pi/2$, corresponding to the top-left corner in figure~\ref{fig:general}. In this case spin-independent interactions are completely absent at tree-level, making loop effects particularly important. Indeed, for the masses and couplings considered in figure~\ref{fig:general} this scenario is not excluded by the bounds from XENON1T but can be tested with LZ. However, the loop contributions depend sensitively on the strength of the couplings, which enter quadratically into the Wilson coefficients. In order to fully assess the importance of loop effects, it is therefore important to consider alternative constraints on the couplings $g_\text{SM}$, $g_\chi$ and $\lambda_{ah}$.

For given values of $m_a$, $m_\chi$ and $g_\text{SM}$ we can fix $g_\chi$ by the requirement that the observed DM relic abundance can be explained in terms of thermal freeze-out via the annihilation processes $\chi \bar{\chi} \to q \bar{q}$ and $\chi \bar{\chi} \to a a$. If the latter process is kinematically allowed, i.e.\ for $m_a < m_\chi$, it will typically give the dominant contribution for $g_\text{SM} \ll 1$, such that the required value for $g_\chi$ becomes independent of $g_\text{SM}$. In this limit, we find $g_\chi \propto m_\chi^{1/2}$ with $g_\chi = 1$ for $m_\chi \approx 500\,\mathrm{GeV}$. For larger $g_\text{SM}$ the calculation becomes more involved and we use \texttt{micrOmegas\_v5.0.6}~\cite{Belanger:2018mqt} to determine the required value for $g_\chi$ numerically.

The coupling of the light spin-0 boson to SM particles can be constrained through a range of flavour physics observables. For $m_a \lesssim m_B \approx 5.2\,\mathrm{GeV}$, these constraints are very strong and effectively exclude the possibility of obtaining observable direct detection signatures~\cite{Dolan:2014ska}. However, almost all of these constraints disappear for larger values of $m_a$. Bounds from radiative $\Upsilon$ decays~\cite{Lees:2012iw,Lees:2011wb} extend to slightly larger masses, but also disappear for $m_a \gtrsim 7 \,\mathrm{GeV}$. Provided the pseudoscalar couples also to leptons (with coupling strength $g_\text{SM} \, m_\ell / v$), another important constraint arises from $B_s \to \mu^+ \mu^-$, which can arise from loop-induced flavour-changing interactions with an off-shell mediator. The resulting branching ratio is given by~\cite{Altmannshofer:2011gn,Batell:2009jf,Dolan:2014ska}
\begin{align}
\frac{\text{BR}(B_s \rightarrow \mu^+ \mu^-)_\text{NP}}{\text{BR}(B_s \rightarrow \mu^+ \mu^-)_\text{SM}} & \simeq  
\frac{g_\text{SM}^4 \, m_t^4 \, m_{B_s}^4}{256\, m_W^4 \, \sin(\theta_W)^4 |C_{10}^\text{SM}|^2 \, \left((m_{B_s}^2-m_a^2)^2+ \Gamma_a^2 \, m_a^2\right)} \log^2\left(\frac{\Lambda^2}{m_t^2}\right)\;,
\end{align}
where $C_{10}^\text{SM} = -4.103$ and $\Lambda$ is the scale of new physics (such as additional charged Higgs bosons needed in a gauge-invariant UV completion). For $\Lambda = 1\,\mathrm{TeV}$ and assuming that $m_a$ is sufficiently far away from $m_{B_s}$, this expression simplifies to
\begin{align}
\frac{\text{BR}(B_s \rightarrow \mu^+ \mu^-)_\text{NP}}{\text{BR}(B_s \rightarrow \mu^+ \mu^-)_\text{SM}} \approx \left( 0.7\,g_\text{SM}\right)^4 \frac{\big(m_{B_s}^2 - (10\,\mathrm{GeV})^2\big)^2}{\big(m_{B_s}^2 - m_a^2\big)^2}\,.
\end{align}
The branching ratio of $B_s\to\mu^+\mu^-$ has been measured with a precision of $20\%$~\cite{Aaij:2017vad} and is found to be in agreement with the SM prediction~\cite{Bobeth:2013uxa}. To obtain an approximate bound on $g_\text{SM}$ we therefore require the new-physics contribution not to exceed 40\% of the SM value. This gives
\begin{equation}
g_\text{SM}\lesssim 1.2\, \sqrt{\frac{\lvert m_a^2 - m_{B_s}^2 \rvert}{(10\,\mathrm{GeV})^2 - m_{B_s}^2 }}\;.
\end{equation}
In other words, even for spin-0 bosons as light as $10\,\mathrm{GeV}$ the coupling strength $g_\text{SM}$ can be of order unity. Constraints of comparable strength have been derived from LHCb dark photon searches within a di-muon channel, see refs.~\cite{Haisch:2018kqx, Aaij:2017rft}.\footnote{For $\phi_\text{SM}$ different from $\pi/2$ there would be additional constraints from observables sensitive to CP-violation, in particular electric dipole moments of leptons~\cite{Chen:2015vqy,Marciano:2016yhf}, nuclei~\cite{Mantry:2014zsa} and atoms~\cite{Stadnik:2017hpa,Dzuba:2018anu}. However, for $\phi_\text{SM} \approx \pi/2$ the spin-0 mediator behaves like a pure pseudoscalar in all observables involving only SM particles, such that these constraints are absent.}

The situation is quite different for the coupling $\lambda_{ah}$ between $a$ and the SM Higgs boson. This coupling induces the decay $h \to aa$ with partial width~\cite{Beniwal:2015sdl}
\begin{equation}
\Gamma_{h\to aa} = \frac{\lambda_{ah}^2 \, v^2}{32\pi \, m_h}\left(1 - \frac{4\,m_a^2}{m_h^2}\right)^{1/2} \; .
\end{equation}
The presence of this decay mode gives rise to exotic Higgs decays and leads to a suppression of the Higgs signal strength in the conventional channels. While the former provide a promising strategy for future searches~\cite{Haisch:2018kqx}, at present the strongest constraints come from a global fit of the measured properties of the SM-like Higgs boson at ATLAS and CMS~\cite{Khachatryan:2016vau}. These fits imply $\text{BR}(h\to aa) < 0.34$, corresponding to $\Gamma_{h\to aa} \lesssim 2\,\mathrm{MeV}$, when simultaneously allowing for modifications of the Higgs boson production cross section, or $\text{BR}(h\to aa) < 0.13$, corresponding to $\Gamma_{h\to aa} \lesssim 0.6\,\mathrm{MeV}$, when assuming the production cross section to be given by the SM prediction. For $m_a \ll m_h/2$, these bounds translate to $\lambda_{ah} \lesssim 0.02$ and $\lambda_{ah} \lesssim 0.01$, respectively. We will conservatively show the weaker bound in the following.

\begin{figure}[t]
  \centering
  \includegraphics[width=0.49\textwidth]{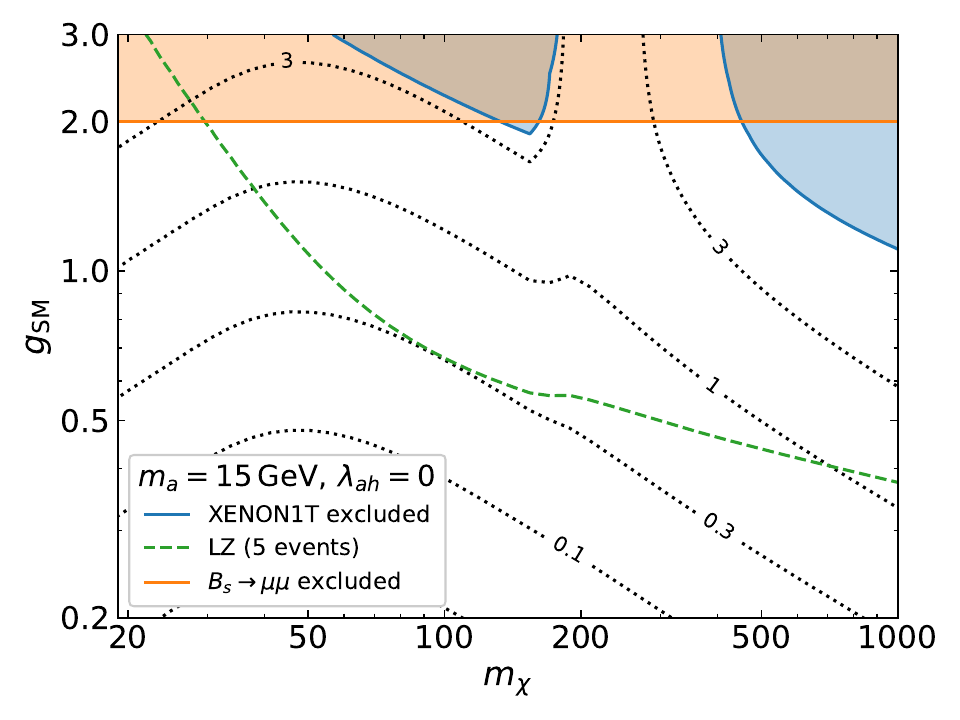}
  \includegraphics[width=0.49\textwidth]{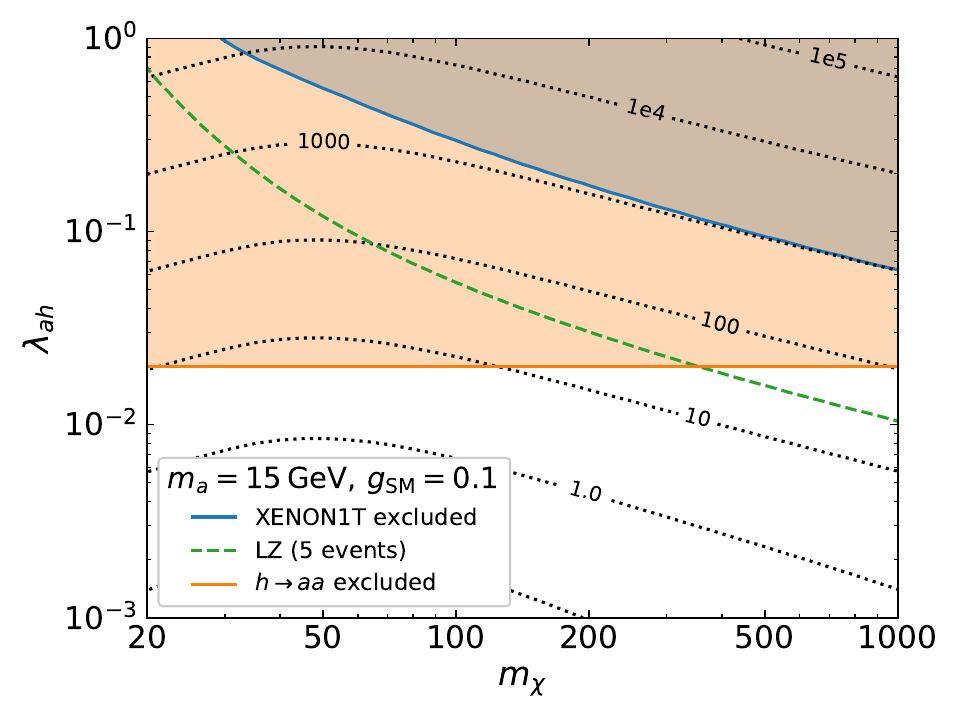}
  \caption{
  Constraints on $g_\text{SM}$ (left) and $\lambda_{ah}$ (right) as a function of $m_\chi$ in a model with maximal CP violation ($\phi_\chi = 0$, $\phi_\text{SM} = \pi/2$). At each point the coupling $g_\chi$ is fixed in such a way  that the observed DM relic abundance is reproduced. The dotted lines indicate the ratio of the predicted number of events in LZ from loop-induced spin-independent interactions and from tree-level momentum-suppressed interactions.
  }
  \label{fig:maxCPV}
\end{figure}

Figure~\ref{fig:maxCPV} summarises the constraints on $g_\text{SM}$ (left) and $\lambda_{ah}$ (right) as a function of $m_\chi$. At each point in the two plots $g_\chi$ is determined by the relic density requirement and we have set $m_a = 15\,\mathrm{GeV}$. Again the solid blue region is excluded by XENON1T and the parameter points for which 5 events are predicted in LZ are indicated by the dashed green line. Dotted black lines in the left panel indicate the ratio of loop-induced spin-independent interactions and tree-level momentum suppressed interactions in terms of the number of predicted events in LZ. As expected, the importance of loop effects grows with increasing $g_\text{SM}$ and with increasing $m_\chi$, corresponding to increasing $g_\chi$. The kinks for $m_\chi \approx 175\,\mathrm{GeV}$ result from the fact that for larger DM masses annihilation into top quarks becomes kinematically allowed and provides an efficient annihilation channel, reducing the required value of $g_\chi$. Furthermore, we observe that loop effects also increase in importance for smaller $m_\chi$. This is related to cancellations occurring in the SD tree-level rate since the meson poles from $\bar{q}i \gamma_5 q$ cancel against those from $G^a_{\mu\nu}\tilde{G}^{a\mu\nu}$ in the DM-neutron coupling for small enough momentum-transfer, i.e.~sufficiently small $m_\chi$.\footnote{Note that the amount of cancellation depends on the values of the $\Delta q$ form factors, for which we use the values given in refs.~\cite{Bishara:2017pfq,Bishara:2017nnn}. Also, higher order contributions to the pseudoscalar and CP-odd gluon form factors may be relevant when cancellations occur, which have however not been computed so far.}

Since direct constraints on $g_\text{SM}$ are quite weak, we find large regions of parameter space where the model can be discovered by LZ. If the interactions of DM arise dominantly from $\lambda_{ah}$, on the other hand, the strong constraints from Higgs measurements imply that there remains only a small region of allowed parameter space that can be explored with LZ. We note that the $h\to a a$ constraint in the right panel is completely independent of $\phi_\chi$ and would hence also apply to a pure pseudoscalar.

\begin{figure}[t]
  \centering
  \includegraphics[width=0.65\textwidth]{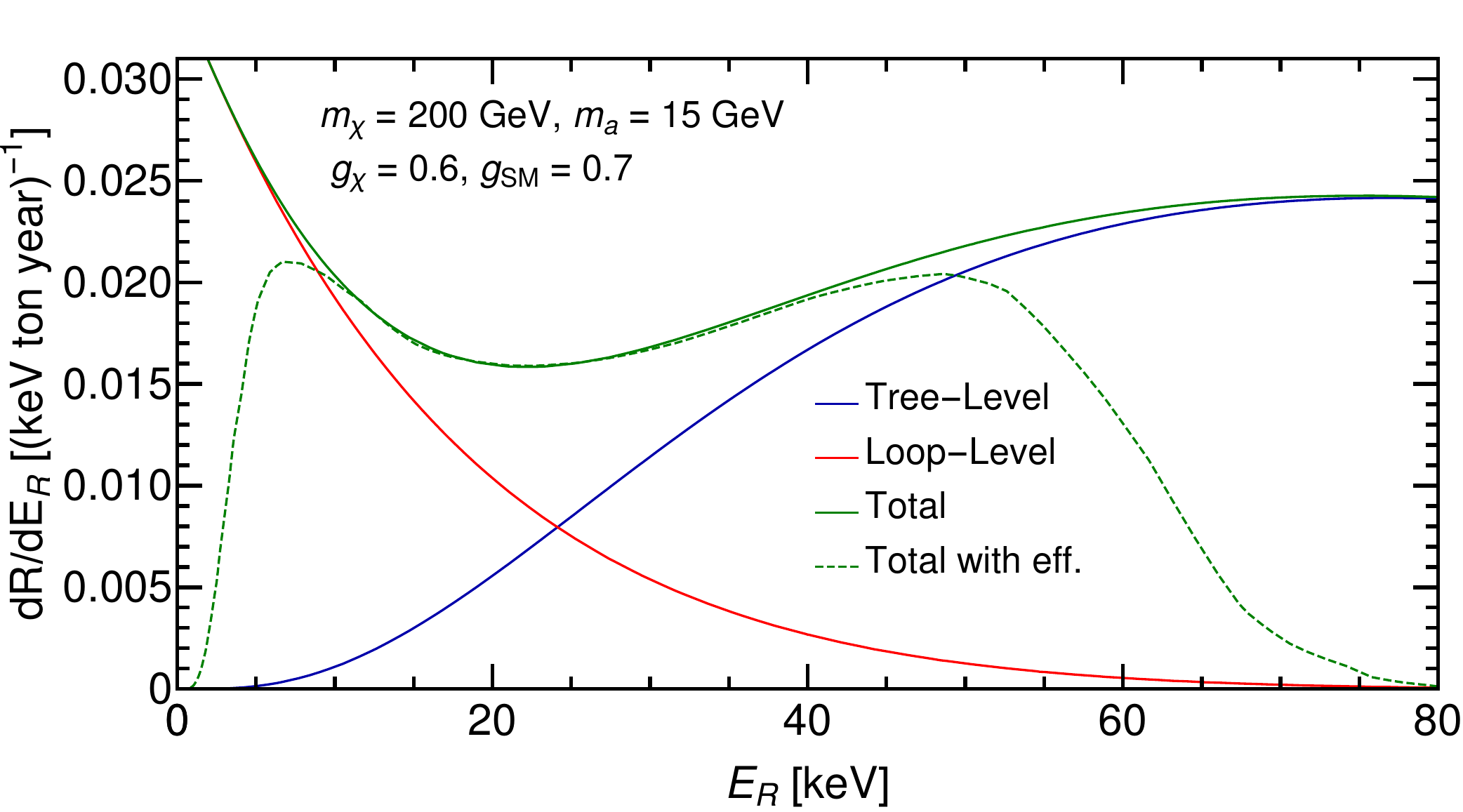}
  \caption{
  Predicted differential event rate in LZ for a specific parameter point in the model with maximal CP violation ($\phi_\chi = 0$, $\phi_\text{SM} = \pi/2$) consistent with all current constraints.}
  \label{fig:dRdE}
\end{figure}

For parameter points close to the XENON1T exclusion bound in the left panel loop effects give a sizeable contribution to the total event rate in direct detection experiments. This observation is illustrated further in figure~\ref{fig:dRdE}, which compares the predicted differential event rates at tree-level and loop-level in LZ for $m_\chi = 200\,\mathrm{GeV}$, $m_a = 15\,\mathrm{GeV}$ and $g_\text{SM} = 0.7$, corresponding to $g_\chi = 0.6$. The tree-level interactions are momentum-suppressed and therefore vanish in the limit $E_\mathrm{R} \to 0$, leading to a maximum around several tens of keV. The differential event rate from loop-induced spin-independent interactions, on the other hand, decreases monotonically with increasing recoil energy. Intriguingly, the two contributions conspire to give a total event rate that is approximately constant across the entire search region. Such a spectrum cannot be obtained from any single non-relativistic operator and could therefore, given enough statistics, be used to identify models like the one discussed here.

A similar interplay between tree level and loop level can arise for $\phi_\chi = \pi/2$, $\phi_\text{SM} = 0$, in which case the tree-level process is coherently enhanced but suppressed by a factor $m_N/m_\chi$ in $c_{11}$, see eq.~(\ref{eq:NRcoeff}). The two scenarios however differ in their dependence on the target material. In particular, if tree-level scattering is spin-dependent, it will be absent in target materials with no nuclear spin, leading to a monotonically falling recoil spectrum from loop-induced spin-independent interactions.

\begin{figure}[t]
\includegraphics[width=7cm]{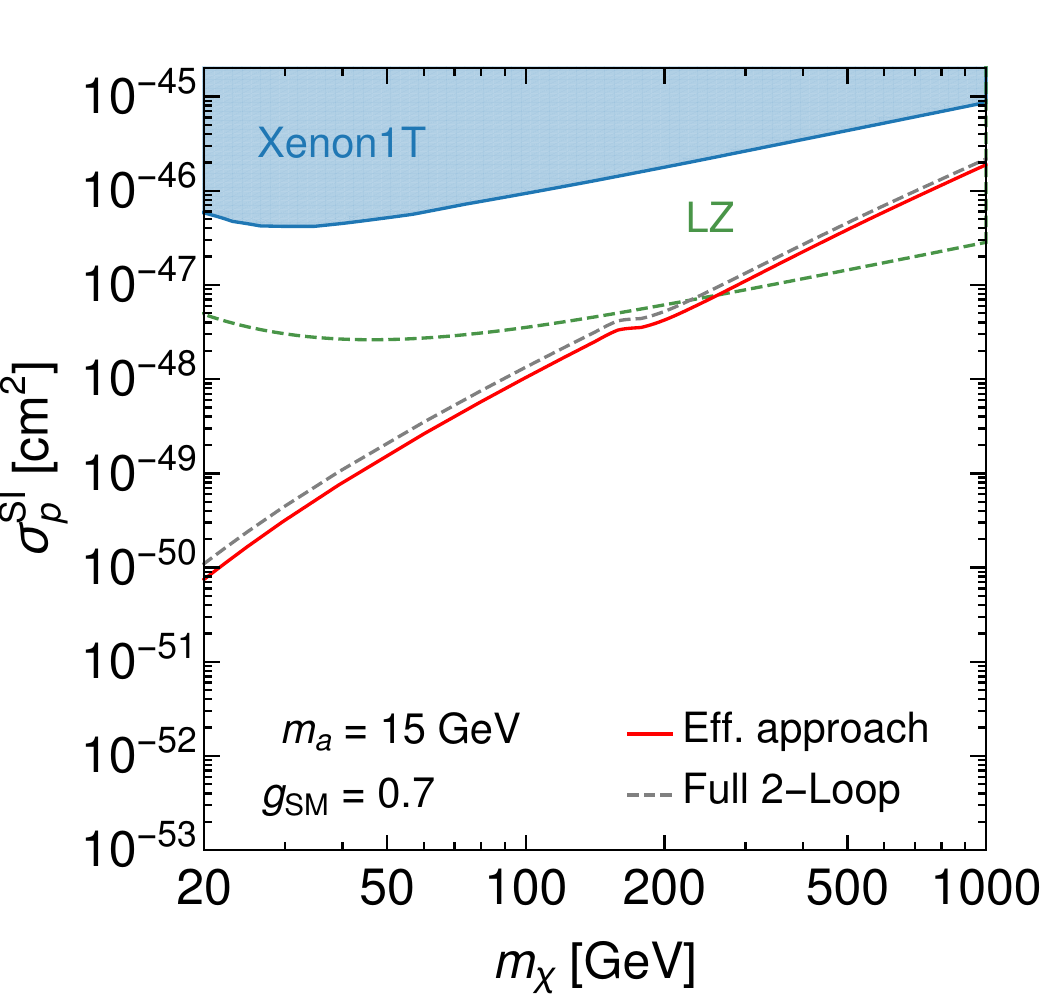}
\hspace{1cm}
\includegraphics[width=7cm]{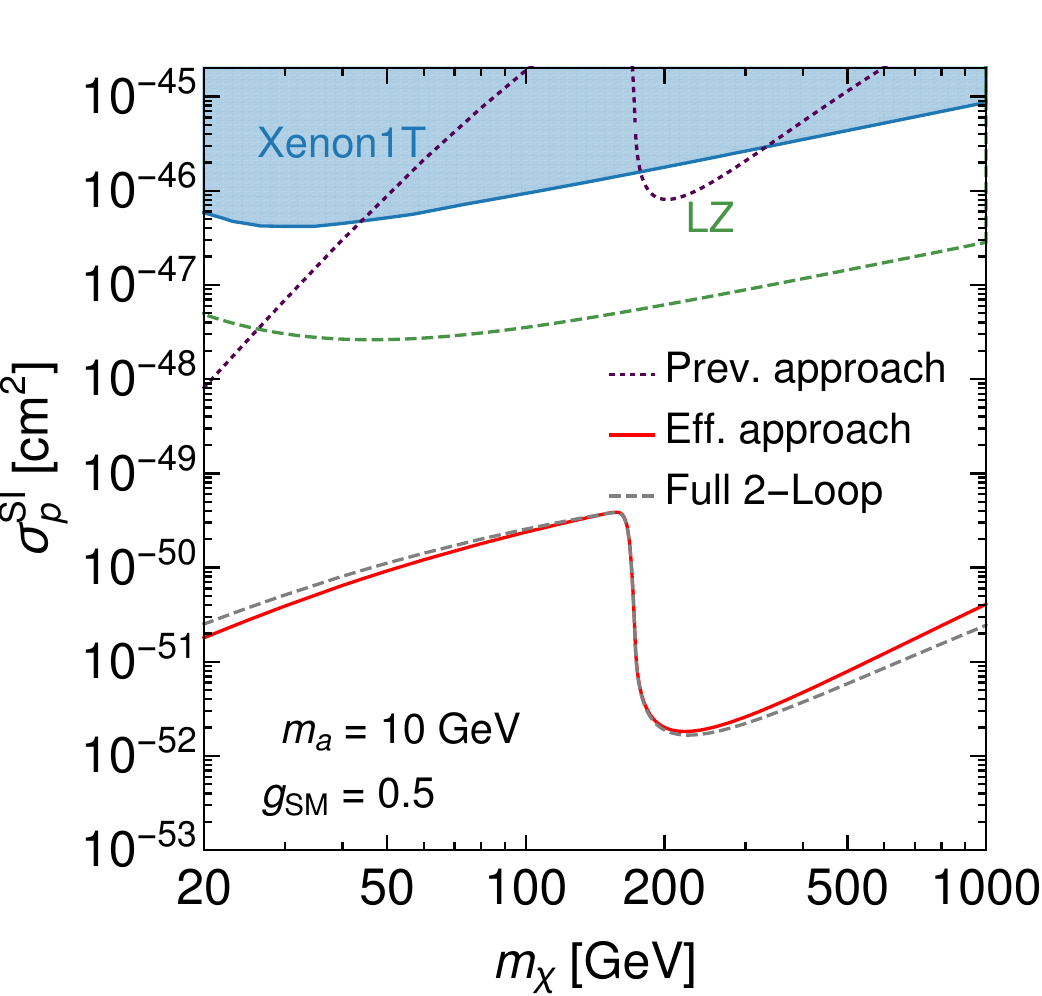}
\caption{Comparison of the effective approach and the full two-loop result for two benchmark points in $m_\chi-\sigma^\text{SI}_{p}\,-$\,plane. In the left panel we consider maximal CP violation with $\phi_\text{SM} = \pi/2$ and $\phi_\chi = 0$, whereas in the right panel we fix $\phi_\text{SM} = \phi_\chi = \pi/2$, i.e.\ pure pseudoscalar phases, and also show the curve corresponding to previous calculational approaches of the two-loop diagram. In both panels $\lambda_{ah}$ is set to zero and $g_\chi$ is fixed such that the correct relic density is reproduced. Note that additional contributions to the differential event rate from momentum-dependent interactions at tree-level may lead to stronger exclusion limits than the ones shown in this plot.}\label{fig:CPComparisonCross}
\end{figure}

Let us finally revisit the discussion of how to approximate two-loop effects in our model. We compare in figure \ref{fig:CPComparisonCross} the spin-independent scattering cross section obtained with our approach (outlined in section~\ref{subsec:eff2Loop}) with the result of a full two-loop calculation including all heavy quarks. The left panel corresponds to the case of maximal CP violation ($\phi_\chi = 0$, $\phi_\text{SM} = \pi/2$), the right panel corresponds to the pure pseudoscalar case ($\phi_\chi = \phi_\text{SM} = \pi/2$). In both cases we fix $g_\chi$ by the relic density requirement and set $g_\text{SM} < 1$, consistent with the bounds discussed above (which are independent of $\phi_\chi$). As can be observed, the top-quark threshold results in a clear feature in the right panel because the underlying annihilation channel $\bar{\chi} \chi \to \bar{f} f$ is s-wave and dominates over $\bar{\chi} \chi \to a a $ for $m_\chi \geq m_t$. In the left panel the former annihilation channel suffers from p-wave suppression~\cite{Berlin:2014tja} and therefore, once the top-quark channel is kinematically allowed, only a rather mild feature is obtained. Importantly, we find very good agreement between the two two-loop approaches, confirming our approach for integrating out top quarks and neglecting the contribution from bottom and charm quarks. In the right panel we also show the cross section obtained if the pseudoscalar is integrated out before all heavy quarks, as previously suggested in refs.~\cite{Arcadi:2017wqi,Sanderson:2018lmj}.\footnote{Here we have used the coefficient $C_{S,q}$ from ref.~\cite{Arcadi:2017wqi} for the top quark and have fixed the overall sign following ref.~\cite{Abe:2018emu}} As pointed out previously~\cite{Abe:2018emu}, this approach leads to a vast overestimation of the loop contribution.

\subsection{CP-violating Higgs portal}\label{subsec:CPHiggs}

As a final example for the importance of loop-effects we consider the fermionic Higgs portal model~\cite{Beniwal:2015sdl,Athron:2018hpc,Abe:2019wku}: 
\begin{equation}\label{eq:LagrCPVHiggs}
 \mathcal{L} = \mathcal{L}_\text{SM} + \overline{\chi} (i \slashed{\partial} - \mu) \chi - \frac{\lambda_{h\chi}}{\Lambda} \left(\cos\psi\, \overline{\chi}\chi + \sin\psi \, \overline{\chi}i\gamma_5 \chi \right) H^\dagger H \; ,
\end{equation}
where $H$ denotes the SM Higgs doublet and $\Lambda$ parametrises the unknown scale of new physics. At first sight, this Lagrangian bears little resemblance to the simplified model discussed so far. After electroweak symmetry breaking, however, the following interactions are generated:
\begin{equation}
 \mathcal{L} \supset - \lambda \, v \, h^3 - \frac{h}{v} \sum_{q} m_q \overline{q} q - \frac{\lambda_{h\chi} \, v}{\Lambda} h \left(\cos\phi \, \overline{\chi} \chi + \sin\phi \, \overline{\chi} i\gamma_5 \chi \right) \, ,
  \label{eq:Lag_psi_pEWSB}
\end{equation}
where $\lambda$ denotes the quartic Higgs self-coupling and
\begin{equation}
\cos\phi = \frac{\mu}{m_\chi} \left(\cos\psi + \frac{1}{2}\frac{\lambda_{h\chi}}{\Lambda} \frac{v^2}{\mu} \right)\;,
\end{equation}
with
\begin{align}
m_\chi &= \sqrt{\left(\mu + \frac{1}{2}\frac{\lambda_{h\chi}}{\Lambda} v^2 \cos\psi \right)^2 + \left(\frac{1}{2}\frac{\lambda_{h\chi}}{\Lambda}v^2 \sin\psi \right)^2} \, .
\end{align}
We can therefore directly apply all the results from section~\ref{sec:loops} with the replacements
\begin{align}\label{eq:Higgs-replace}
 m_a = m_h, \quad g_\chi = \frac{\lambda_{h\chi} v}{\Lambda}, \quad g_\text{SM} = 1, \quad \phi_\text{SM} = 0, \quad \phi_\chi = \phi, \quad \lambda_{ah} = - 6 \lambda = - 3 \frac{m_h^2}{v^2}\; .
\end{align}
The factor of 6 in the last expression is necessary to ensure that the correct Feynman rule is obtained in spite of different combinatorial factors. The free parameters of this model are hence $m_\chi$, $\lambda_{h\chi}/\Lambda$ and $\phi$. 

Note that, since we study loop processes within this model, one should in principle include all operators involving DM particles and SM fields that contribute at the order of $1/\Lambda^2$, in particular dimension six operators coupling the DM vector and axial-vector current to the corresponding SM quark currents. Here we implicitly assume the absence of new spin-1 particles at the high energy \mbox{scale $\Lambda$} that would induce such operators. Operators including scalar, pseudoscalar or tensor couplings between DM and quarks would generally be accompanied with a factor of $m_q$ and would therefore only contribute at higher order, i.e.~$1/\Lambda^3$.

\begin{figure}[t]
  \centering
  \includegraphics[width=0.5\textwidth]{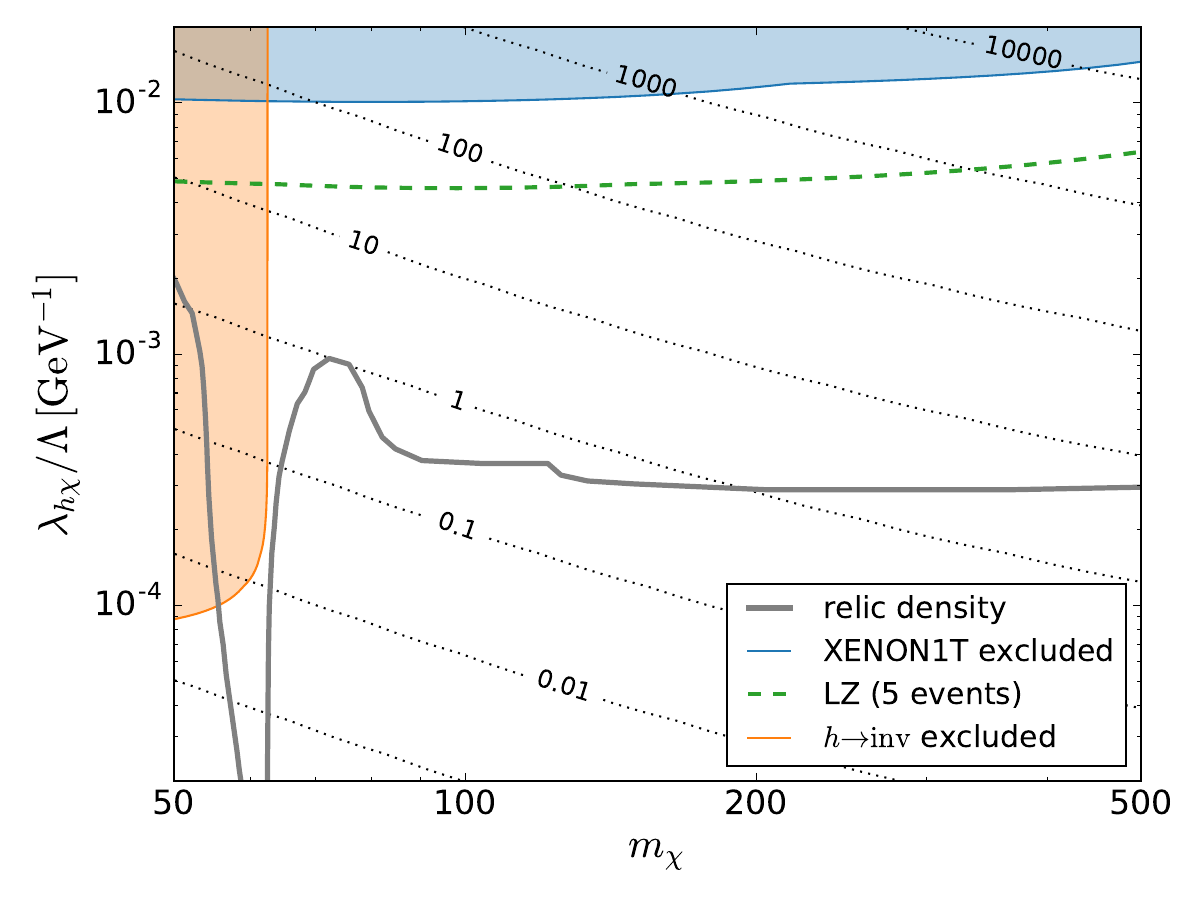}
  \caption{
  Constraints and preferred parameter regions for the CP-violating Higgs portal model with $\phi = \pi/2$. The dotted lines indicate the ratio of loop-induced spin-independent interactions and tree-level momentum-suppressed interactions in terms of the predicted number of events in LZ.
  }
  \label{fig:HP}
\end{figure}

For $\phi \neq 0$ the model violates CP and spin-independent scattering is suppressed proportional to $\cos^2 \phi$. As $\phi$ approaches $\pi/2$, loop effects are therefore expected to become increasingly important. We confirm this expectation in figure~\ref{fig:HP}, which shows constraints on $\lambda_{h\chi}/\Lambda$ as a function of $m_\chi$. Dotted lines indicate the ratio of loop-induced spin-independent interactions to tree-level momentum-suppressed interactions (in terms of the expected number of events in LZ). In the parameter range that can be probed by direct detection experiments, this ratio is significantly larger than unity, implying that the sensitivity of direct detection experiments stems almost exclusively from loop-induced interactions.\footnote{We note that our effective description of top-quark loops overestimates the contribution to the Wilson coefficient for spin-independent scattering by up to a factor of 3 compared to the full two-loop result. However, by far the dominant contribution to this coefficient arises from triangle diagrams, making the difference between the effective description and the full two-loop calculation irrelevant.}

In figure~\ref{fig:HP} we also indicate the parameter regions excluded by the constraint $\text{BR}(h \to \text{inv}) < 0.26$~\cite{Aaboud:2018sfi,Sirunyan:2018owy} as well as the combinations of $\lambda_{h\chi}/\Lambda$ and $m_\chi$ for which the observed DM relic abundance can be reproduced via annihilations into SM particles~\cite{Beniwal:2015sdl}. The requirement of EFT validity, $\lambda_{h\chi}/\Lambda < 2\pi/m_\chi$~\cite{Athron:2018hpc}, is satisfied in the entire parameter region shown in figure~\ref{fig:HP}. We find that for $\phi = \pi/2$ constraints from direct detection experiments are rather weak and only probe parameter regions where the standard freeze-out calculation predicts $\chi$ to be a sub-dominant DM component. For these parameter regions we implicitly assume that the abundance of $\chi$ is set by a non-standard mechanism (e.g.\ a particle-antiparticle asymmetry) such that $\chi$ accounts for all of the DM. If, on the other hand, bounds from direct detection experiments are rescaled based on the abundance of $\chi$ obtained from standard freeze-out, as done e.g.~in ref.~\cite{Athron:2018hpc}, loop-induced direct detection signals do not provide relevant constraints on the CP-violating Higgs portal model for the foreseeable future.

We point out that within this model two additional loop diagrams with one insertion each of $h \bar{\chi} i \gamma_5 \chi$ and $h^2 \bar{\chi} i \gamma_5 \chi$ contribute to the amplitudes relevant for direct detection. These diagrams are however UV divergent, which indicates a dependence on the specific UV completion of the effective Higgs portal operator. Replacing the UV divergence with $\log{(\Lambda^2/m_\chi^2)}$ and setting $\Lambda = 1\,\mathrm{TeV}$, we find that these loops can be numerically important and increase the predicted event rates (see appendix~\ref{app:uv-div-higgs} for additional details). Nevertheless, these additional contributions are still not large enough for near-future direct detection experiments to reach the relic density line shown in figure~\ref{fig:HP}. To make this statement more precise would require the choice of a specific UV completion.

\section{Conclusions}
\label{sec:conclusions}

Future direct detection experiments will reach such a high level of sensitivity to the interactions between DM and quarks that loop effects become increasingly important. This is particularly true in models where tree-level scattering is suppressed, such that loop-induced interactions may give the dominant contribution and yield potentially observable signals. In the present work we have studied such a set-up in the context of a  spin-0 particle $a$ mediating the interaction between DM and SM fermions. In contrast to previous studies, we allow general CP phases and therefore cover scalar, pseudoscalar and CP-violating interactions. Moreover, we include a trilinear coupling between $a$ and the SM Higgs boson which generally arises in UV completions of this model and can have important phenomenological consequences.

For certain combinations of CP phases standard spin-independent contributions are strongly suppressed or even fully absent at tree-level, such that a proper calculation of the interactions induced at loop-level is crucial. In our model, these arise from Higgs-induced triangle diagrams, box diagrams for light quarks (both shown in figure~\ref{fig:CPDiagram1}) as well as the two-loop process involving heavy quarks shown in figure~\ref{fig:CPDiagram4}. In particular the two-loop process gives an important contribution, which is difficult to estimate without performing the full calculation. To address this challenge, we have presented a novel approach for simplifying the two-loop calculation significantly for heavy quark masses (schematically illustrated in figure~\ref{fig:CPDiagram4}). Provided the top quark gives the dominant contribution and the mediator is light compared to the top quark, this approach makes it possible to circumvent the two-loop calculation entirely and obtain an accurate estimate that is much easier to calculate and implement. A comparison between the two approaches is provided in figure~\ref{fig:CPComparisonCoeff}.

As illustrated in figure~\ref{fig:general}, loop effects are most important when at least one of the CP phases is close to $\pi/2$ (corresponding to pseudoscalar interactions). Moreover, they gain in importance as the sensitivity of direct detection experiments improves. A particularly interesting observation is that the recoil rates induced at tree- and loop-level can be comparable, resulting in a roughly constant event rate over the whole energy window (see figure~\ref{fig:dRdE}). Since such a spectrum cannot be generated from a single type of interaction, it will be very interesting to perform a detailed statistical analysis of how to discriminate the model studied here from alternative hypotheses.

Finally, we have studied the impact of spin-independent loop-induced interactions on the CP-violating fermionic Higgs portal model. Our results show that loop-level effects allow future direct detection experiments to probe parameter regions that would be otherwise inaccessible. Nevertheless, loop-level contributions are still too small to enable direct detection experiments to reach the parameter regions preferred by thermal freeze-out if the CP phase is close to $\pi/2$.

Based on the results presented in this work, we conclude that a general spin-0 mediator offers an interesting possibility to evade current direct detection bounds even with $\mathcal{O}(1)$ couplings while still maintaining promising detection prospects for future years. It will therefore be important to investigate how such a simplified model can arise from a more complete theory, such as an extended Higgs sector with spontaneous CP breaking. Such an embedding will provide new insights on the relations between the different couplings and allow for a more accurate analysis of the constraints from flavour physics and precision observables.

\acknowledgments

We thank Giorgio Arcadi and Sebastian Wild for discussions and Joachim Brod for valuable comments on the manuscript. This  work  is  funded  by  the  Deutsche Forschungsgemeinschaft (DFG) through the Emmy Noether Grant No.\ KA 4662/1-1 and the Collaborative Research Center TRR 257 ``Particle Physics Phenomenology after the Higgs Discovery''.

\newpage

\appendix

\section{One-loop Wilson coefficients}
\label{app:one-loop}

In this appendix we provide details on the one-loop calculations relevant for section~\ref{sec:one-loop}.

\subsection{Loop functions}\label{app:one-loopfcts}
We define the Passarino-Veltman functions $C_i$ that appear in our one-loop calculations according to the standard notation~\cite{Passarino:1978jh}
\begin{align}
\int \frac{\text{d}^{4}k}{(2\pi)^4} \frac{1}{[(p+k)^2 - M^2]\, [k^2 - m^2]^2} &= \frac{i}{(4\pi)^2}\, C_0(p^2,\, m^2,\,M^2)\;,\\[0.75em]
\int \frac{\text{d}^{4}k}{(2\pi)^4} \frac{k^\mu}{[(p+k)^2 - M^2]\, [k^2 - m^2]^2} &= \frac{i}{(4\pi)^2}\,p^\mu\,C_2(p^2,\, m^2,\,M^2)\;.
\end{align}
The $X_2$ and $Y_2$ functions are given by~\cite{Abe:2015rja}
\begin{align}
\int \frac{\text{d}^{4}k}{(2\pi)^4} \frac{1}{[(p+k)^2 - M^2]\,k^2\,[k^2 - m^2]^2} &= \frac{i}{(4\pi)^2}\,X_2(p^2,\,M^2,\,0,\,m^2)\;,\\[0.75em]
\int \frac{\text{d}^{4}k}{(2\pi)^4} \frac{k^\mu}{[(p+k)^2 - M^2]\,k^2\,[k^2 - m^2]^2} &= \frac{i}{(4\pi)^2}\,p^\mu\,Y_2(p^2,\,M^2,\,0,\,m^2)\;,
\end{align}
which will reappear in the full two-loop approach in appendix~\ref{app:two-loopfcts}. Finally, we define the $Z$ functions
\begin{align}
\begin{split}
\int \frac{\text{d}^{4}k}{(2\pi)^4} \frac{k^\mu k^\nu}{[(p+k)^2 - M^2]\,k^4\,[k^2 - m^2]^2} &\\
&\hspace{-5.9cm}= \frac{i}{(4\pi)^2}\Big(p^\mu p^\nu\,Z_{11}(p^2,\,M^2,\,m^2) + g^{\mu \nu}\,Z_{00}(p^2,\,M^2,\,m^2)  \Big)\;,
\end{split}\\ \notag\\
\begin{split}
\int \frac{\text{d}^{4}k}{(2\pi)^4} \frac{k^\mu k^\nu k^\alpha}{[(p+k)^2 - M^2]\,k^4\,[k^2 - m^2]^2}&\\
 =\frac{i}{(4\pi)^2}\Big(p^\mu p^\nu p^\alpha \,Z_{111}(p^2,\,M^2,\,m^2) &+ \big(g^{\mu\nu}p^\alpha +g^{\alpha\mu}p^\nu+g^{\nu\alpha}p^\mu\big)\,Z_{001}(p^2,\,M^2,\,m^2)  \Big)\;.
\end{split}
\raisetag{2.8\normalbaselineskip}
\end{align}
All of these functions can be calculated readily with \texttt{Package-X}~\cite{Patel:2015tea}. As examples, we quote the expression for $C_0(p^2,\,m^2,\,M^2)$,
\begin{align}
\notag
C_0(p^2,\, m^2,\,M^2) = &-\frac{\log \left(\frac{m^2}{M^2}\right)}{2 p^2} + \Bigg[ \frac{\left(m^2-M^2-p^2\right)}{p^2 \sqrt{m^4-2 m^2 M^2-2 m^2 p^2+M^4-2 M^2 p^2+p^4}} \\
&\hspace{-2.5cm}\times \log \left(\frac{m^2+\sqrt{m^4-2 m^2 M^2-2 m^2 p^2+M^4-2 M^2 p^2+p^4}+M^2-p^2}{2 m M}\right)\Bigg]\;,
\end{align}
and $Z_{11}(M^2,\,M^2,\,m^2)$ evaluated using the on-shell condition $p^2 = M^2$,
\begin{align}
\notag
Z_{11}(M^2,\,M^2,\,m^2) =&\,\frac{1}{3 m^2 M^4}-\frac{\log \left(\frac{m^2}{M^2}\right)}{6 M^6}\\
&\hspace{-0.2cm}+\frac{\sqrt{m^2 \left(m^2-4 M^2\right)} \left(m^4-2 m^2 M^2-2 M^4\right) \log \left(\frac{\sqrt{m^2 \left(m^2-4 M^2\right)}+m^2}{2 m M}\right)}{3 m^4 M^6
   \left(m^2-4 M^2\right)}\;.
\end{align}
The remaining coefficients can be computed analogously.

\subsection{Box diagram computation and coefficients}\label{app:one-loopwilson}

\begin{figure}[t]
\centering
\includegraphics[width=5cm]{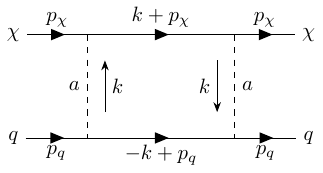}
\hspace{2.5cm}
\includegraphics[width=5cm]{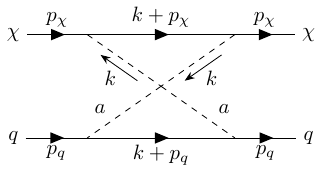}
\caption{Box and corresponding crossed diagram including visualization of the momentum flow in the limit of no momentum transfer.}\label{fig:BoxDiagram}
\end{figure}

\noindent
For the computation of the box and its crossed diagram shown in figure \ref{fig:BoxDiagram} we follow the procedure from ref.~\cite{Abe:2018emu}, which allows us to derive the coefficients for the twist-2 operators. We first start with the amplitude, which can be expressed as
\begin{align}
\notag
&i\mathcal{M}_\text{Box}= g_\chi^2\,g_\text{SM}^2\,\frac{m_q^2}{v^2}\\
\notag
& \hspace{0cm}\times\bar{u}_{\chi}(p_\chi)\int \frac{\text{d}^4 k}{(2\pi)^4} \left(\frac{\slashed{k}}{(k+p_\chi)^2 - m_\chi^2} + \frac{2\, m_\chi\, [\cos^2(\phi_\chi) + i \gamma_5 \sin(\phi_\chi) \cos(\phi_\chi)]}{(k+p_\chi)^2-m_\chi^2}  \right) u_{\chi}(p_\chi)\\
\notag
&\hspace{0cm}\times\frac{1}{(k^2 - m_a^2)^2}\,\bar{u}_{q}(p_q) \left(\frac{-\slashed{k}}{(k-p_q)^2 - m_q^2} + \frac{2\, m_q\, [\cos^2(\phi_\text{SM}) + i \gamma_5 \sin(\phi_\text{SM}) \cos(\phi_\text{SM})]}{(k-p_q)^2-m_q^2}  \right) u_{q}(p_q)\\
&\hspace{0cm}+\text{crossed diagram}\;,
\end{align}
where the crossed diagram is obtained by the replacement $k \rightarrow -k$ within $\bar{u}_{q}(p_q) ... u_{q}(p_q)$ and we have suppressed the sum over the quark species. Now we expand the amplitude in terms of $p_q$, as this is the smallest scale involved in the diagram:
\begin{align}
\frac{-1}{(k-p_q)^2 - m_q^2} + \frac{1}{(k+p_q)^2 - m_q^2} &= \frac{-4\,k\cdot p_q}{k^4} + \mathcal{O}((p_q)^2)\;,\\
\frac{1}{(k-p_q)^2 - m_q^2} + \frac{1}{(k+p_q)^2 - m_q^2} &= \frac{2}{k^2}+ \mathcal{O}((p_q)^2)\;,
\end{align}
where we have employed the on-shell condition $(p_q)^2 = m_q^2$. The amplitude then reads
\begin{align}
\notag
&i\mathcal{M}_\text{Box}= g_\chi^2\,g_\text{SM}^2\,\frac{m_q^2}{v^2} \\
\notag
& \hspace{0cm}\times\bar{u}_{\chi}(p_\chi)\int \frac{\text{d}^4 k}{(2\pi)^4} \left(\frac{\slashed{k}}{(k+p_\chi)^2 - m_\chi^2} + \frac{2 \,m_\chi\, [\cos^2(\phi_\chi) + i \gamma_5 \sin(\phi_\chi) \cos(\phi_\chi)]}{(k+p_\chi)^2-m_\chi^2}  \right) u_{\chi}(p_\chi)\\ 
\notag
& \hspace{0cm}\times\frac{1}{(k^2 - m_a^2)^2}\,\bar{u}_{q}(p_q) \left(\frac{-\slashed{k}\,4\,k\cdot p_q}{k^4} + \frac{4\,m_q\, [\cos^2(\phi_\text{SM})+i \gamma_5 \sin(\phi_\text{SM}) \cos(\phi_\text{SM}) ]}{k^2} \right) u_{q}(p_2)\\
&\hspace{0cm}+ \mathcal{O}((p_q)^2)\;.
\end{align}
We can now identify the loop functions defined in appendix~\ref{app:one-loopfcts}, construct the corresponding effective Lagrangian and use the following decomposition:
\begin{align}
\bar{q} i \partial^\mu \gamma^\nu q = \mathcal{O}^q_{\mu\nu} + \bar{q} \frac{i \partial^\mu \gamma^\nu - i \partial^\nu \gamma^\mu}{2} q +  \frac{1}{4} g^{\mu\nu} m_q \bar{q} q\;.
\end{align}
This then yields the effective box diagram Lagrangian given in eqs.~(\ref{effbox1}) and~(\ref{effbox2}) with the coefficients given by
\begin{align}
\notag
\mathcal{C}^\text{box}_{1,q} =&\,\kappa_q\,m_\chi\Big(-m_\chi^2 \,Z_{111}(m_\chi^2,\,m_\chi^2,\,m_a^2) - 6 \, Z_{001}(m_\chi^2,\,m_\chi^2,\,m_a^2)\\
&+ 4 \,\cos^2(\phi_\text{SM})\, Y_2(m_\chi^2,\,m_\chi^2,\,0,\,m_a^2) -2\, m_\chi^2 \cos^2(\phi_\chi)\,Z_{11}(m_\chi^2,\,m_\chi^2,\,m_a^2)\\
\notag
&-8\, \cos^2(\phi_\chi)\, Z_{00}(m_\chi^2,\,m_\chi^2,\,m_a^2)+ 8\,  \cos^2(\phi_\chi) \cos^2(\phi_\text{SM}) \,X_2(m_\chi^2,\,m_\chi^2,\,0,\,m_a^2)\Big)\;,\\[0.75em]
\notag
\mathcal{C}^\text{box}_{2,q} =& \,\kappa_q\,m_\chi\Big(-2 \,m_\chi^2 \cos(\phi_\chi) \sin(\phi_\chi)\,Z_{11}(m_\chi^2,\,m_\chi^2,\,m_a^2)\\
& -8 \, \cos(\phi_\chi) \sin(\phi_\chi)\, Z_{00}(m_\chi^2,\,m_\chi^2,\,m_a^2)\\
\notag
& + 8 \, \cos(\phi_\chi) \sin(\phi_\chi) \cos^2(\phi_\text{SM})\,X_2(m_\chi^2,\,m_\chi^2,\,0,\,m_a^2)\Big)\;,\\[0.75em]
\begin{split}
\mathcal{C}^\text{box}_{3,q} =&\,\kappa_q\,m_\chi\Big( 4 \, \cos(\phi_\text{SM}) \sin(\phi_\text{SM})\,  Y_2(m_\chi^2,\,m_\chi^2,\,0,\,m_a^2)\\
&+8\, \cos^2(\phi_\chi) \cos(\phi_\text{SM}) \sin(\phi_\text{SM})\, X_2(m_\chi^2,\,m_\chi^2,\,0,\,m_a^2)\Big)\;,
\end{split}\\[0.75em]
\mathcal{C}^\text{box}_{4,q} =&  \,8\,\kappa_q\, m_\chi \cos(\phi_\chi) \sin(\phi_\chi) \cos(\phi_\text{SM}) \sin(\phi_\text{SM})\, X_2(m_\chi^2,\,m_\chi^2,\,0,\,m_a^2)\;,\\[0.75em]
\mathcal{C}^\text{box}_{5,q} =& -8\,\kappa_q\, Z_{001}(m_\chi^2,\,m_\chi^2,\,m_a^2)\;,\\[0.75em]
\mathcal{C}^\text{box}_{6,q} =&\,\kappa_q\,m_\chi\Big(-4\,Z_{111}(m_\chi^2,\,m_\chi^2,\,m_a^2) - 8 \, \cos^2(\phi_\chi)\, Z_{11}(m_\chi^2,\,m_\chi^2,\,m_a^2)\Big)\;,\\[0.75em]
\mathcal{C}^\text{box}_{7,q} =&-8\,\kappa_q\, m_\chi \cos(\phi_\chi) \sin(\phi_\chi)\,Z_{11}(m_\chi^2,\,m_\chi^2,\,m_a^2)\;,
\end{align}
where we have used the shorthand notation 
\begin{align}
\label{eq:kappaq}
\kappa_q = \frac{g^2_\chi\, g^2_\text{SM}\,m_q^2}{16 v^2 \pi^2}\;.
\end{align}

\section{Details on two-loop calculations}
\label{app:two-loop}

In this appendix we provide details on the two-loop calculations relevant for section~\ref{subsec:eff2Loop}.

\subsection{Loop functions}\label{app:two-loopfcts}
For the two-loop computation presented in appendix \ref{app:two-loopcomp}, we will need further loop functions. The Passarino-Veltman functions $D_i$ read in their standard notation~\cite{Passarino:1978jh}
\begin{align}
\int \frac{\text{d}^{4}k}{(2\pi)^4} \frac{1}{[(p+k)^2 - M^2]\, [k^2 - m^2]^3} &= \frac{i}{(4\pi)^2}\, D_0(p^2,\, m^2,\,M^2)\;,\\[0.75em]
\int \frac{\text{d}^{4}k}{(2\pi)^4} \frac{k^\mu}{[(p+k)^2 - M^2]\, [k^2 - m^2]^3} &= \frac{i}{(4\pi)^2}\,p^\mu\,D_3(p^2,\, m^2,\,M^2)\;.
\end{align}
We further define the $X_n$ and $Y_n$ functions by~\cite{Abe:2015rja}
\begin{align}
\int \frac{\text{d}^{4}k}{(2\pi)^4} \frac{1}{[k^2 - \frac{m_q^2}{x(1-x)}]^n\,[(p_\chi+k)^2 - m_\chi^2]\,[k^2 - m_a^2]} &= \frac{i}{(4\pi)^2}\,X_n\left(p_\chi^2,\,m_\chi^2,\,m_a^2,\,\tfrac{m_q^2}{x(1-x)}\right)\;,\\[0.75em]
\int \frac{\text{d}^{4}k}{(2\pi)^4} \frac{k^\mu}{[k^2 - \frac{m_q^2}{x(1-x)}]^n\,[(p_\chi+k)^2 - m_\chi^2]\,[k^2 - m_a^2]} &= \frac{i}{(4\pi)^2}\,p_\chi^\mu\,Y_n\left(p_\chi^2,\,m_\chi^2,\,m_a^2,\,\tfrac{m_q^2}{x(1-x)}\right)\;.
\end{align}
Using partial fraction decomposition, we can derive the following relations for $X_n$
\begin{align}
\begin{split}
X_1&\left(p_\chi^2,\,m_\chi^2,\,m_a^2,\,\tfrac{m_q^2}{x(1-x)}\right) \\
&= \frac{1}{m_a^2 - \frac{m_q^2}{x(1-x)}} \left(B_0\left(p_\chi^2,\,m_a^2,\,m_\chi^2\right) - B_0\left(p_\chi^2,\,\tfrac{m_q^2}{x(1-x)},\,m_\chi^2\right) \right)\;,
\end{split}\\ \notag\\
\begin{split}
X_2&\left(p_\chi^2,\,m_\chi^2,\,m_a^2,\,\tfrac{m_q^2}{x(1-x)}\right) \\
&= \frac{1}{m_a^2 - \frac{m_q^2}{x(1-x)}} \left(X_1\left(p_\chi^2,\,m_\chi^2,\,m_a^2,\,\tfrac{m_q^2}{x(1-x)}\right) - C_0\left(p_\chi^2,\,\tfrac{m_q^2}{x(1-x)},\,m_\chi^2\right) \right)\;,
\end{split}\\ \notag \\
\begin{split}
X_3&\left(p_\chi^2,\,m_\chi^2,\,m_a^2,\,\tfrac{m_q^2}{x(1-x)}\right) \\
&= \frac{1}{m_a^2 - \frac{m_q^2}{x(1-x)}} \left(X_2\left(p_\chi^2,\,m_\chi^2,\,m_a^2,\,\tfrac{m_q^2}{x(1-x)}\right) - D_0\left(p_\chi^2,\,\tfrac{m_q^2}{x(1-x)},\,m_\chi^2\right) \right)\;,
\end{split}
\end{align}
as well as for $Y_n$~\cite{Abe:2018emu} 
\begin{align}
\begin{split}
Y_1&\left(p_\chi^2,\,m_\chi^2,\,m_a^2,\,\tfrac{m_q^2}{x(1-x)}\right) \\
&= \frac{1}{m_a^2 - \frac{m_q^2}{x(1-x)}} \left(B_1\left(p_\chi^2,\,m_a^2,\,m_\chi^2\right) - B_1\left(p_\chi^2,\,\tfrac{m_q^2}{x(1-x)},\,m_\chi^2\right) \right)\;,
\end{split}\\ \notag\\
\begin{split}
Y_2&\left(p_\chi^2,\,m_\chi^2,\,m_a^2,\,\tfrac{m_q^2}{x(1-x)}\right) \\
&= \frac{1}{m_a^2 - \frac{m_q^2}{x(1-x)}} \left(Y_1\left(p_\chi^2,\,m_\chi^2,\,m_a^2,\,\tfrac{m_q^2}{x(1-x)}\right) - C_2\left(p_\chi^2,\,\tfrac{m_q^2}{x(1-x)},\,m_\chi^2\right) \right)\;,
\end{split}\\ \notag \\
\begin{split}
Y_3&\left(p_\chi^2,\,m_\chi^2,\,m_a^2,\,\tfrac{m_q^2}{x(1-x)}\right) \\
&= \frac{1}{m_a^2 - \frac{m_q^2}{x(1-x)}} \left(Y_2\left(p_\chi^2,\,m_\chi^2,\,m_a^2,\,\tfrac{m_q^2}{x(1-x)}\right) - D_3\left(p_\chi^2,\,\tfrac{m_q^2}{x(1-x)},\,m_\chi^2\right) \right)\;,
\end{split}
\end{align}
where the loop functions denoted by $B_0(p^2,\,m^2,\,M^2)$ and $B_1(p^2,\,m^2,\,M^2)$ are implemented in \texttt{LoopTools}~\cite{Hahn:1998yk}.\footnote{We could in principle evaluate the functions $X_n$ and $Y_n$ directly with \texttt{Package-X}~\cite{Patel:2015tea} and then perform the numerical integration appearing in the two-loop calculation using their explicit expressions. Numerical stability improves, however, when the functions are decomposed as presented here.} We further need the derivatives of these functions with respect to $m_a^2$
\begin{align}
\begin{split}
\frac{\partial}{\partial m_a^2} X_1&\left(p_\chi^2,\,m_\chi^2,\,m_a^2,\,\tfrac{m_q^2}{x(1-x)}\right)\\
&= \frac{1}{m_a^2 - \frac{m_q^2}{x(1-x)}}  \left(C_0(p_\chi^2,\,m_a^2,\, m_\chi^2) - X_1\left(p_\chi^2,\,m_\chi^2,\,m_a^2,\,\tfrac{m_q^2}{x(1-x)}\right)\right)\;,
\end{split}\\ \notag\\
\notag
\frac{\partial}{\partial m_a^2} X_2&\left(p_\chi^2,\,m_\chi^2,\,m_a^2,\,\tfrac{m_q^2}{x(1-x)}\right)\\
&= \frac{1}{m_a^2 - \frac{m_q^2}{x(1-x)}}  \left(\frac{\partial}{\partial m_a^2} X_1\left(p_\chi^2,\,m_\chi^2,\,m_a^2,\,\tfrac{m_q^2}{x(1-x)}\right) - X_2\left(p_\chi^2,\,m_\chi^2,\,m_a^2,\,\tfrac{m_q^2}{x(1-x)}\right)\right)\;,\\ \notag\\
\notag
\frac{\partial}{\partial m_a^2} X_3&\left(p_\chi^2,\,m_\chi^2,\,m_a^2,\,\tfrac{m_q^2}{x(1-x)}\right)\\
&= \frac{1}{m_a^2 - \frac{m_q^2}{x(1-x)}}  \left(\frac{\partial}{\partial m_a^2} X_2\left(p_\chi^2,\,m_\chi^2,\,m_a^2,\,\tfrac{m_q^2}{x(1-x)}\right) - X_3\left(p_\chi^2,\,m_\chi^2,\,m_a^2,\,\tfrac{m_q^2}{x(1-x)}\right)\right)\;,
\end{align}
as well as~\cite{Abe:2018emu}
\begin{align}
\begin{split}
\frac{\partial}{\partial m_a^2} Y_1&\left(p_\chi^2,\,m_\chi^2,\,m_a^2,\,\tfrac{m_q^2}{x(1-x)}\right)\\
&= \frac{1}{m_a^2 - \frac{m_q^2}{x(1-x)}}  \left(C_2(p_\chi^2,\,m_a^2,\, m_\chi^2) - Y_1\left(p_\chi^2,\,m_\chi^2,\,m_a^2,\,\tfrac{m_q^2}{x(1-x)}\right)\right)\;,
\end{split}\\ \notag\\
\notag
\frac{\partial}{\partial m_a^2} Y_2&\left(p_\chi^2,\,m_\chi^2,\,m_a^2,\,\tfrac{m_q^2}{x(1-x)}\right)\\
&= \frac{1}{m_a^2 - \frac{m_q^2}{x(1-x)}}  \left(\frac{\partial}{\partial m_a^2} Y_1\left(p_\chi^2,\,m_\chi^2,\,m_a^2,\,\tfrac{m_q^2}{x(1-x)}\right) - Y_2\left(p_\chi^2,\,m_\chi^2,\,m_a^2,\,\tfrac{m_q^2}{x(1-x)}\right)\right)\;,\\ \notag\\
\notag
\frac{\partial}{\partial m_a^2} Y_3&\left(p_\chi^2,\,m_\chi^2,\,m_a^2,\,\tfrac{m_q^2}{x(1-x)}\right)\\
&= \frac{1}{m_a^2 - \frac{m_q^2}{x(1-x)}}  \left(\frac{\partial}{\partial m_a^2} Y_2\left(p_\chi^2,\,m_\chi^2,\,m_a^2,\,\tfrac{m_q^2}{x(1-x)}\right) - Y_3\left(p_\chi^2,\,m_\chi^2,\,m_a^2,\,\tfrac{m_q^2}{x(1-x)}\right)\right)\;.
\end{align}

\subsection{Review of the two-loop computation}\label{app:two-loopcomp}

The two-loop computation was recently presented for a purely pseudoscalar mediator in the context of a 2HDM~\cite{Abe:2018emu} (see ref.~\cite{Hill:2014yka} for a related computation in a different context). For our model, however, we have to generalise the results to arbitrary CP phases in the SM and dark sector. We use the opportunity to present intermediate calculational steps of the derivation not explicitly shown in ref.~\cite{Abe:2018emu}. Our starting point is the calculation of the leading order effective vertices between the spin-0 \mbox{mediator $a$} and gluons, which we treat as background. To simplify the computation we employ the Fock-Schwinger gauge, in which the gluon field can directly be expressed in terms of the field strength~\cite{Hisano:2010ct,Novikov:1983gd}.
\begin{figure}[t]
\begin{subfigure}{0.3\textwidth}
\includegraphics[width=4cm]{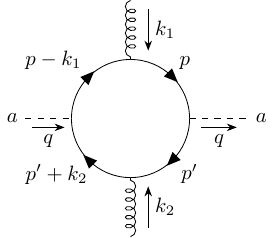}
\end{subfigure}
\hspace*{\fill}
\begin{subfigure}{0.32\textwidth}
\includegraphics[width=\linewidth]{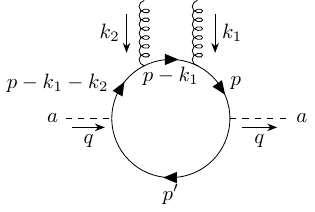}
\end{subfigure}
\hspace*{\fill}
\begin{subfigure}{0.32\textwidth}
\includegraphics[width=\linewidth]{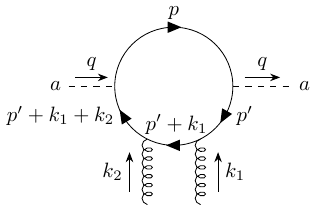}
\end{subfigure}
\caption{Visualization of the first step in the two-loop calculation where we compute the heavy quark loop contribution to the effective vertices between $a$ and gluons~\cite{Novikov:1983gd}. Here $p'$ is given by $p' = p - q$.}\label{fig:app2Loop:1}
\end{figure}

The general amplitude can thus be written as
\begin{align}
\begin{split}
\label{GeneralFSAmplaaGG}
i \mathcal{M}^{aaG}_\text{2-Loop}= - &\sum_{q=c,b,t} \left( \frac{i g_\text{SM} \,m_q}{v} \right)^2\\ 
&\hspace{-2.4cm}\times \int \frac{\text{d}^{4}p}{(2\pi)^4} \text{Tr}^\text{Dirac}_\text{Colour}\bigg[\big[\cos(\phi_\text{SM}) + i \gamma_5\sin(\phi_\text{SM})\big]\, i S(p)\, \big[\cos(\phi_\text{SM}) + i \gamma_5 \sin(\phi_\text{SM})\big]\, i \tilde{S}(p-q)\,\bigg]\;,
\end{split}
\raisetag{3.3\normalbaselineskip}
\end{align}
where two different coloured fermion propagators, $S(p)$ and $\tilde{S}(p-q)$, occur, as the Fock-Schwinger gauge breaks translational invariance. Since we are interested in the effective vertices $a a\,G^{a}_{\mu\nu} G^{a\mu\nu}$ and $a a\,G^{a}_{\mu\nu} \widetilde{G}^{a\mu\nu}$, we have to consider the following terms in the propagators
\begin{align}
\notag
i S(p) =&\,i S^{(0)}(p) +\int \text{d}^{4}k_1\,i S^{(0)}(p)\,\, g_s \gamma^\alpha \frac{1}{2} G_{\mu\alpha} \bigg(\frac{\partial}{\partial k_{1\mu}} \delta^{(4)}(k_1)\bigg)\,\,i S^{(0)}(p-k_1)\\
&+ \int \text{d}^{4}k_1 \text{d}^{4}k_2\, i S^{(0)}(p)\,\,g_s \gamma^\alpha \frac{1}{2} G_{\mu\alpha} \bigg(\frac{\partial}{\partial k_{1\mu}} \delta^{(4)}(k_1)\bigg)\,\,i S^{(0)}(p-k_1)\\
\notag
&\hspace{0.5cm}\times g_s \gamma^\beta \frac{1}{2} G_{\nu\beta} \bigg(\frac{\partial}{\partial k_{2\nu}} \delta^{(4)}(k_2)\bigg)\,\,i S^{(0)}(p-k_1 - k_2) + ...\;,
\end{align}
\begin{align}
\notag
i \tilde{S}(p) =& \,i S^{(0)}(p) + \int \text{d}^{4}k_1\,i S^{(0)}(p+k_1)\,\, g_s \gamma^\beta \frac{1}{2} G_{\nu\beta} \bigg(\frac{\partial}{\partial k_{1\nu}}\delta^{(4)}(k_1)\bigg) \,\,i S^{(0)}(p)\\
&+\int \text{d}^{4}k_1 \text{d}^{4}k_2\,i S^{(0)}(p+k_1 + k_2)\,\,g_s \gamma^\alpha \frac{1}{2} G_{\mu\alpha} \bigg(\frac{\partial}{\partial k_{2\mu}}\delta^{(4)}(k_2)\bigg)\\
\notag
&\hspace{0.5cm}\times i S^{(0)}(p+k_1)\,\,g_s \gamma^\beta \frac{1}{2} G_{\nu\beta} \bigg(\frac{\partial}{\partial k_{1\nu}}\delta^{(4)}(k_1)\bigg)\,\,i S^{(0)}(p)+...\;,
\end{align}
where terms with derivatives acting on $G_{\mu\nu}$ are neglected since they are not relevant for the present work. Further we have used $i S^{(0)}(p) = i (\slashed{p} + m)/(p^2 - m^2)$ and $G^{\mu\nu} = G^{a\mu\nu}\,t^a$ with $t^a$ being an $SU(3)$ generator fulfilling $\text{Tr}[t^a t^b] = \delta^{ab}/2$. Inserting these expressions into eq.~(\ref{GeneralFSAmplaaGG}), we identify three terms relevant for the computation of vertices involving two gluon field strength tensors. These terms are visualised in figure~\ref{fig:app2Loop:1}. We will present the various calculational steps for the term involving a gluon field strength tensor each from $S(p)$ and $\tilde{S}(p-q)$, illustrated in the left panel in figure~\ref{fig:app2Loop:1}. The corresponding term reads
\begin{align}
\notag
i \mathcal{M}^{aaG}_\text{2-Loop}\supset - &\sum_{q=c,b,t} \left( \frac{ig_\text{SM}\,m_q}{v} \right)^2\\ 
\notag
&\hspace{-2.5cm}\times\frac{g_s^2}{8}\,G^{a}_{\alpha\mu} G^{a}_{\beta\nu} \int \frac{\text{d}^{4}p}{(2\pi)^4} \frac{\partial}{\partial k_{1\mu}} \frac{\partial}{\partial k_{2\nu}}\text{Tr}^\text{Dirac}\bigg[\big[\cos(\phi_\text{SM}) + i \gamma_5\sin(\phi_\text{SM})\big] \frac{\slashed{p}+m}{p^2 - m^2} \gamma^\alpha \frac{\slashed{p} - \slashed{k}_1+m}{(p-k_1)^2 - m^2}\\
&\hspace{-1.5cm}\times \big[\cos(\phi_\text{SM}) + i \gamma_5\sin(\phi_\text{SM})\big] \frac{\slashed{p} - \slashed{q} + \slashed{k}_2+m}{(p-q+k_2)^2 - m^2} \gamma^\beta \frac{\slashed{p} - \slashed{q}+m}{(p-q)^2 - m^2} \bigg]_{k_1 = k_2 = 0}\;,
\end{align}
where we performed the trace over colour indices as well as partial integration regarding $k_1$ and $k_2$ and switched the indices of the gluon field strengths. After performing the derivatives with respect to $k_{1\mu}$ and $k_{2\nu}$, setting $k_1 = k_2 = 0$ afterwards and evaluating the trace using \texttt{Package-X}~\cite{Patel:2015tea}, we now have to project out the leading spin-independent and spin-dependent term. The spin-independent term can be obtained by rewriting\footnote{As pointed out in ref.~\cite{Berlin:2015njh}, the signs in front of $\mathcal{O}^{G}_{\beta\mu}$ and $\mathcal{O}^{G}_{\alpha\nu}$ differ from eq.~(50) in ref.~\cite{Hisano:2010ct}. As we are not including the terms containing the twist-2 operator, this difference will not play a role.}
\begin{align}
\begin{split}
\label{GGdecompFull}
G^{a}_{\alpha\mu} G^{a}_{\beta\nu} &= \frac{1}{12} G^{a}_{\rho\sigma} G^{a\rho\sigma} (g_{\alpha\beta}g_{\mu\nu} - g_{\alpha\nu} g_{\beta\mu})+ \frac{1}{2} g_{\alpha \beta}\, \mathcal{O}^{G}_{\mu\nu}\\
& \quad+ \frac{1}{2} g_{\mu \nu} \,\mathcal{O}^{G}_{\alpha\beta} - \frac{1}{2} g_{\alpha \nu} \,\mathcal{O}^{G}_{\beta\mu} - \frac{1}{2} g_{\beta \mu} \,\mathcal{O}^{G}_{\alpha\nu}+ \mathcal{O}^{G}_{\alpha\mu\beta\nu}\;,
\end{split}
\end{align}
where we have introduced the twist-2 gluon operator $ \mathcal{O}^{G}_{\mu\nu}$ and a higher spin \mbox{operator $\mathcal{O}^{G}_{\alpha\mu\beta\nu}$}
\begin{align}
\mathcal{O}^{G}_{\mu\nu} &= G^{a\rho}_{\hphantom{a\rho}\mu} G^a_{\hphantom{a}\rho\nu} - \frac{1}{4} g_{\mu\nu} G^{a}_{\rho\sigma} G^{a\rho\sigma}\;,\\
\begin{split}
\mathcal{O}^{G}_{\alpha\mu\beta\nu}&= G^{a}_{\alpha\mu} G^{a}_{\beta\nu} - \frac{1}{2} g_{\alpha\beta}G^{a\rho}_{\hphantom{a\rho}\mu} G^a_{\hphantom{a}\rho\nu} -  \frac{1}{2} g_{\mu\nu}G^{a\rho}_{\hphantom{a\rho}\alpha} G^a_{\hphantom{a}\rho\beta}\\
&\quad+\frac{1}{2} g_{\alpha\nu}G^{a\rho}_{\hphantom{a\rho}\beta} G^a_{\hphantom{a}\rho\mu} + \frac{1}{2} g_{\beta\mu}G^{a\rho}_{\hphantom{a\rho}\alpha} G^a_{\hphantom{a}\rho\nu}\\
&\quad+\frac{1}{6} G^{a}_{\rho\sigma} G^{a\rho\sigma}(g_{\alpha\beta}g_{\mu\nu} - g_{\alpha\nu} g_{\beta\mu})\;.
\end{split}
\end{align}
These operators do not contribute to $G^{a}_{\rho\sigma} G^{a\rho\sigma}$ and give sub-leading SI interactions such that they are neglected in the present work. We therefore only have to consider the replacement
\begin{align}
\label{GGdecompEff}
G^{a}_{\alpha\mu} G^{a}_{\beta\nu} \rightarrow \frac{1}{12} G^{a}_{\rho\sigma} G^{a\rho\sigma} (g_{\alpha\beta}g_{\mu\nu} - g_{\alpha\nu} g_{\beta\mu})\;,
\end{align}
for the spin-independent term. The spin-dependent term involving $G^{a}_{\mu\nu} \widetilde{G}^{a\mu\nu}$ can be obtained straightforwardly by isolating the term in the trace containing the $\epsilon$-tensor. Putting both contributions together, we then obtain
\begin{align}
\notag
i \mathcal{M}^{aaG}_\text{2-Loop}\supset - \sum_{q=c,b,t} \bigg( &\frac{i g_\text{SM} \,m_q}{v} \bigg)^2 \Bigg\{\frac{g_s^2}{96}\, G^{a}_{\rho\sigma} G^{a\rho\sigma}\int \frac{\text{d}^{4}p}{(2\pi)^4} \,\frac{-48\,[2 \,m_q^2\, \cos(2\phi_\text{SM}) + p\cdot (p-q)]}{[p^2-m_q^2]^2\,[(p-q)^2-m_q^2]^2}\\
&+ g_s^2 \,G^{a}_{\rho\sigma} \widetilde{G}^{a\rho\sigma} \int \frac{\text{d}^{4}p}{(2\pi)^4}  \frac{m_q^2\, \sin(2\phi_\text{SM})}{[p^2-m_q^2]^2\,[(p-q)^2-m_q^2]^2}\Bigg\}\;.
\end{align}
Since the momentum $q$ corresponds to the second loop momentum of the full two-loop diagram it is not helpful to simply perform the loop integral over $p$, as one has to perform a second loop integral afterwards. It is more advantageous to use a Feynman parameter $x$ instead:
\begin{gather}
\frac{1}{A^2 \,B^2} = \int_0^1 \text{d}x\, \frac{6\, x(1-x)}{(xA+(1-x)B)^4}\;,\\
x((p-q)^2-m^2) + (1-x)(p^2-m^2) = (p-qx)^2-m^2-q^2x(-1+x)\;.
\end{gather}
Shifting $p \rightarrow p+qx$ and performing the loop integral over $p$, we obtain finally
\begin{align}
\notag
i \mathcal{M}^{aaG}_\text{2-Loop}\supset& \,i \sum_{q=c,b,t} \left( \frac{g_\text{SM}\, m_q}{v}\right)^2 \frac{g^2_s}{32\pi^2}\, G^{a}_{\rho\sigma} G^{a\rho\sigma} \\
\notag
&\hspace{1.25cm}\times \int_0^1 \text{d}x \,\Bigg\{\frac{-2\,m_q^2\,[\cos(2\phi_\text{SM}) - \frac{1}{2}]\,x(1-x)}{[m_q^2 - q^2 x(1-x)]^2} + \frac{x(1-x)}{[m_q^2 - q^2 x(1-x)]}\Bigg\}\\[1em]
&\hspace{0.1cm}+ i \sum_{q=c,b,t} \left( \frac{g_\text{SM}\,m_q}{v}\right)^2 \frac{g^2_s}{16\pi^2}\, G^{a}_{\rho\sigma}  {\widetilde{G}}^{a\rho\sigma} \int_0^1 \text{d}x\, \frac{m_q^2\, \sin(2\phi_\text{SM})\, x(1-x)}{[m_q^2 - q^2 x(1-x)]^2}\;.
\end{align}
One can proceed in a similar fashion for the remaining two terms in eq.~(\ref{GeneralFSAmplaaGG}), which contribute equally to the amplitude, such that the full amplitude reads
\begin{align}
\label{app:aaG2Loop}
i \mathcal{M}^{aaG}_\text{2-Loop} = &\,i\,d^\text{\,full}_{G}(q^2) \frac{\alpha_s}{12\pi}\, G^{a}_{\rho\sigma} G^{a\rho\sigma} + i\,d^\text{\,full}_{\widetilde{G}}(q^2) \frac{\alpha_s}{8\pi} \, G^{a}_{\rho\sigma} \widetilde{G}^{a\rho\sigma}\;,
\end{align}
with
\begin{align}
\begin{split}
\label{app:dGFull}
d^\text{\,full}_{G}(q^2) = &\,\sum_{q=c,b,t} \left( \frac{g_\text{SM}\, m_q}{v}\right)^2  \int_0^1 \text{d}x\, \Bigg\{\frac{\frac{3}{2}\, x(1-x)}{[m_q^2 - q^2 x(1-x)]}\\
&\hspace{-1.75cm}+\frac{m_q^2}{2}\,\frac{3\,(1-x)x + 2\,(-1-x+x^2)\,\cos(2\phi_\text{SM})}{[m_q^2 - q^2 x(1-x)]^2} - m_q^4\, \frac{1-3x+3x^2-(1-x)x\,\cos(2\phi_\text{SM})}{[m_q^2 - q^2 x(1-x)]^3}\Bigg\}\;,
\end{split}\raisetag{3.6\normalbaselineskip}\\ \notag\\
\label{app:dGdualFull}
d^\text{\,full}_{\widetilde{G}}(q^2) = & \,\sum_{q=c,b,t} \left( \frac{g_\text{SM}\, m_q}{v}\right)^2  \int_0^1 \text{d}x\, \frac{m_q^2\, \sin(2\phi_\text{SM})}{[m_q^2 - q^2 x(1-x)]^2}\;,
\end{align}
where we have made the integral over the Feynman parameter symmetric under $x \rightarrow 1-x$. This result agrees with eq.~(B.73) from ref.~\cite{Abe:2018emu} for $\phi_\text{SM} = \pi/2$. Moreover, when performing a heavy quark expansion on eq.~(\ref{app:aaG2Loop}), we recover the effective Lagrangian from eq.~(\ref{effaaGMatch}). \\
It is now straightforward to perform the integration from the remaining triangle diagram of the full two-loop approach, visualised after the first arrow in figure~\ref{fig:CPDiagram4}, for which the amplitude can be written as
\begin{align}
\begin{split}
i \mathcal{M}^\text{SI}_{\text{2-Loop}} = - \bar{u}_\chi (p_\chi)\, g_\chi^2 \int& \frac{\text{d}^{4}q}{(2\pi)^4}\, \frac{\slashed{q} + 2 \,m_\chi\, [\cos^2(\phi_\chi) + i \gamma_5 \sin(\phi_\chi) \cos(\phi_\chi)]}{[(p_\chi + q)^2 - m_\chi^2]\,[q^2-m_a^2]^2}\, u_\chi(p_\chi)\\
&\times \bigg[d^\text{\,full}_{G}(q^2) \,\frac{\alpha_s}{12\pi}\, G^{a}_{\rho\sigma} G^{a\rho\sigma} + d^\text{\,full}_{\widetilde{G}}(q^2)\, \frac{\alpha_s}{8\pi} \, G^{a}_{\rho\sigma} \widetilde{G}^{a\rho\sigma}  \bigg]\;.
\end{split}
\end{align}
In terms of the various loop functions defined in appendix~\ref{app:two-loopfcts}, we can map this amplitude onto
\begin{align}
\mathcal{L}^\text{SI}_{\text{2-Loop}} &= \left(\mathcal{C}^\text{full}_{G,S}\, \bar{\chi} \chi+\mathcal{C}^\text{full}_{G,PS}\, \bar{\chi} i \gamma_5 \chi\,\right)\frac{-\alpha_s}{12\pi}\,G^{a}_{\mu\nu} G^{a\mu\nu}\;,\\
\mathcal{L}^\text{SD}_{\text{2-Loop}} &= \left(\mathcal{C}^\text{full}_{\widetilde{G},S}\, \bar{\chi} \chi+\mathcal{C}^\text{full}_{\widetilde{G},PS}\, \bar{\chi} i \gamma_5 \chi\, \right)\frac{\alpha_s}{8\pi}\,G^{a}_{\mu\nu} {\widetilde{G}}^{a\mu\nu}\;,
\end{align}
with
\begin{align}
\begin{split}
\mathcal{C}^\text{full}_{G,S} = \frac{1}{(4\pi)^2} \sum_{q=c,b,t} &\left( \frac{g_\chi \,g_\text{SM}\, m_q}{v}\right)^2 \\
&\hspace{-0.6cm}\times\bigg\{ m_\chi \,F_Y(p_\chi^2,\,m_\chi^2,\,m_a^2,\,m_q^2) + 2 \,m_\chi \cos^2(\phi_\chi)\,F_X(p_\chi^2,\,m_\chi^2,\,m_a^2,\,m_q^2)\bigg\}\;,
\end{split}\raisetag{3.3\normalbaselineskip}\\ \notag\\
\mathcal{C}^\text{full}_{G,PS} = \frac{1}{(4\pi)^2} \sum_{q=c,b,t} &\left( \frac{g_\chi\,g_\text{SM}\, m_q}{v}\right)^2 2\, m_\chi \sin(\phi_\chi) \cos(\phi_\chi)\,F_X(p_\chi^2,\,m_\chi^2,\,m_a^2,\,m_q^2)\;,\\ \notag \\
\begin{split}
\mathcal{C}^\text{full}_{\widetilde{G},S} = -\frac{1}{(4\pi)^2} \sum_{q=c,b,t} &\left( \frac{g_\chi\,g_\text{SM}\, m_q}{v}\right)^2 \\
&\hspace{-0.6cm}\times\bigg\{ m_\chi\,E_Y(p_\chi^2,\,m_\chi^2,\,m_a^2,\,m_q^2) + 2 \,m_\chi \cos^2(\phi_\chi)\,E_X(p_\chi^2,\,m_\chi^2,\,m_a^2,\,m_q^2)\bigg\}\;,
\end{split}\raisetag{3.3\normalbaselineskip}\\ \notag\\
\mathcal{C}^\text{full}_{\widetilde{G},PS} = -\frac{1}{(4\pi)^2} \sum_{q=c,b,t} &\left( \frac{g_\chi\,g_\text{SM}\, m_q}{v}\right)^2 2\, m_\chi \sin(\phi_\chi) \cos(\phi_\chi)\, E_X(p_\chi^2,\,m_\chi^2,\,m_a^2,\,m_q^2)\;.
\end{align}
Here we have introduced the shorthand notation
\begin{align}
\notag
F_\Lambda(p_\chi^2,\,m_\chi^2,\,m_a^2,\,m_q^2) &= \int_0^1 \text{d}x \bigg[ -\frac{3}{2}\,\frac{\partial}{\partial m_a^2}\, \Lambda_1\left(p_\chi^2,\,m_\chi^2,\,m_a^2,\,\tfrac{m_q^2}{x(1-x)}\right)\\[0.75em]
\notag
&\hspace{-2.5cm}+\frac{m_q^2}{2}\, \frac{3\,(1-x)x + 2\,(-1-x+x^2)\,\cos(2\phi_\text{SM})}{x^2 (1-x)^2}\frac{\partial}{\partial m_a^2} \,\Lambda_2\left(p_\chi^2,\,m_\chi^2,\,m_a^2,\,\tfrac{m_q^2}{x(1-x)}\right)\\[0.75em]
&\hspace{-2.5cm}+ m_q^4\, \frac{1-3x+3x^2-(1-x)x\,\cos(2\phi_\text{SM})}{x^3(1-x)^3} \frac{\partial}{\partial m_a^2}\, \Lambda_3\left(p_\chi^2,\,m_\chi^2,\,m_a^2,\,\tfrac{m_q^2}{x(1-x)}\right)\bigg]\;,
\end{align}
\begin{align}
E_\Lambda(p_\chi^2,\,m_\chi^2,\,m_a^2,\,m_q^2) &= \int_0^1 \text{d}x\,\bigg[m_q^2\, \frac{\sin(2\phi_\text{SM})}{x^2(1-x)^2}\, \frac{\partial}{\partial m_a^2} \,\Lambda_2\left(p_\chi^2,\,m_\chi^2,\,m_a^2,\,\tfrac{m_q^2}{x(1-x)}\right)\bigg]\;,
\end{align}
with $\Lambda = X,Y$.

\section{Nuclear form factors}\label{app:rel-nuc-wilson}
In this appendix we define the nuclear form factors required to calculate the effective interactions between DM and nucleons. For the spin-independent interactions we need the following nuclear form factors~\cite{Shifman:1978zn,Jungman:1995df}:
\begin{align}
\label{defFormFactorSI}
q &=u,d,s:&\quad\langle N|m_q \bar{q} q|N\rangle &= m_N f^N_{q}\;,\\
Q &=c,b,t:& \quad \langle N|m_Q \bar{Q} Q|N\rangle &= \langle N|-\frac{\alpha_s}{12\pi} G^{a}_{\mu\nu}G^{a\mu\nu}|N\rangle = \frac{2}{27}m_N f^N_{G}\;,\\
q&=u,d,s,c,b:&\quad \langle N| \mathcal{O}^q_{\mu\nu}|N\rangle &= \frac{1}{m_N} \Big(p^N_\mu p^N_\nu - \frac{1}{4} m_N^2 g_{\mu\nu}\Big) \Big(q^{N}(2) + \bar{q}^{N}(2)\Big)\;,
\end{align}
where $f^N_{q}$ and $f^N_G$ are form factors, $m_N$ is the nucleon mass, $q^{N}(2)$ and $\bar{q}^{N}(2)$ are the second moments of the quark parton distribution functions and $p^N_\mu$ is the nucleon four-momentum. The values of the form factors for light quarks are taken from \texttt{micrOmegas}~\cite{Belanger:2018mqt}\footnote{We refer to ref.~\cite{Ellis:2018dmb} for a discussion of the uncertainties of these form factors, in particular regarding the strange quark matrix element.}
\begin{align}
f^p_{u} &= 0.01513\;,\qquad& f^p_{d} &= 0.0191\;,\qquad&f^p_{s} &= 0.0447\;,\\
f^n_{u} &= 0.0110\;,\qquad&f^n_{d} &= 0.0273\;,\qquad&f^n_{s} &= 0.0447\;,
\end{align}
which are related to the gluon form factor via~\cite{Jungman:1995df}
\begin{align}
\hspace{-0.5cm}f^p_{G} = 1 - \sum_{q=u,d,s} f^p_{q} = 0.92107\;,\qquad
f^n_{G} = 1 - \sum_{q=u,d,s} f^n_{q} = 0.917\;.
\end{align}
The second moments are calculated at the scale $\mu = m_Z$ by using CTEQ PDFs~\cite{Abe:2018emu,Pumplin:2002vw}. For the proton, one finds
\begin{align}
u^p(2) &= 0.22\;,\qquad&\bar{u}^p(2) &= 0.034\;,\\
d^p(2) &= 0.11\;,\qquad&\bar{d}^p(2) &= 0.036\;,\\
s^p(2) &= 0.026\;,\qquad&\bar{s}^p(2) &= 0.026\;,\\
c^p(2) &= 0.019\;,\qquad&\bar{c}^p(2) &= 0.019\;,\\
b^p(2) &= 0.012\;,\qquad&\bar{b}^p(2) &= 0.012\;,
\end{align}
whereas for the neutron the up- and down-quark values have to be interchanged. 
 
For the spin-dependent interactions we need the following form factors:
\begin{align}
q &=u,d,s:&\quad\langle N'|m_q \bar{q} i \gamma_5 q|N\rangle &= F^{q/N}_P(q^2)\;,\\
Q &=c,b,t:& \quad \langle N'|m_Q\bar{Q}i \gamma_5 Q|N\rangle &= \langle N'|\frac{\alpha_s}{8\pi} G^{a}_{\mu\nu}{\widetilde{G}}^{a\mu\nu}|N\rangle = F^N_{\widetilde{G}}(q^2)\; ,
\end{align}
where $N'$ refers to a change of nucleon momentum. This explicit dependence on the momentum transfer $q^\mu$ arises from non-negligible $\pi$ and $\eta$ pole contributions. The corresponding expressions are given in eqs.~(A30) and~(A42) of ref.~\cite{Bishara:2017pfq} and are implemented in \texttt{DirectDM}~\cite{Bishara:2017nnn}.

\section{UV-divergent loops in CP-violating Higgs portal}\label{app:uv-div-higgs}

\begin{figure}[t]
\center
\includegraphics[width=4.5cm]{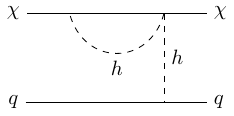}
\hspace{2cm}
\includegraphics[width=4.5cm]{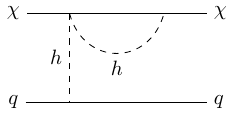}
\caption{Additional loop diagrams within the CP-violating Higgs portal model which also induce purely spin-independent interactions for $\phi = \pi/2$.}\label{fig:UVDivHiggs}
\end{figure}

\begin{figure}[t]
  \centering
  \includegraphics[width=0.5\textwidth]{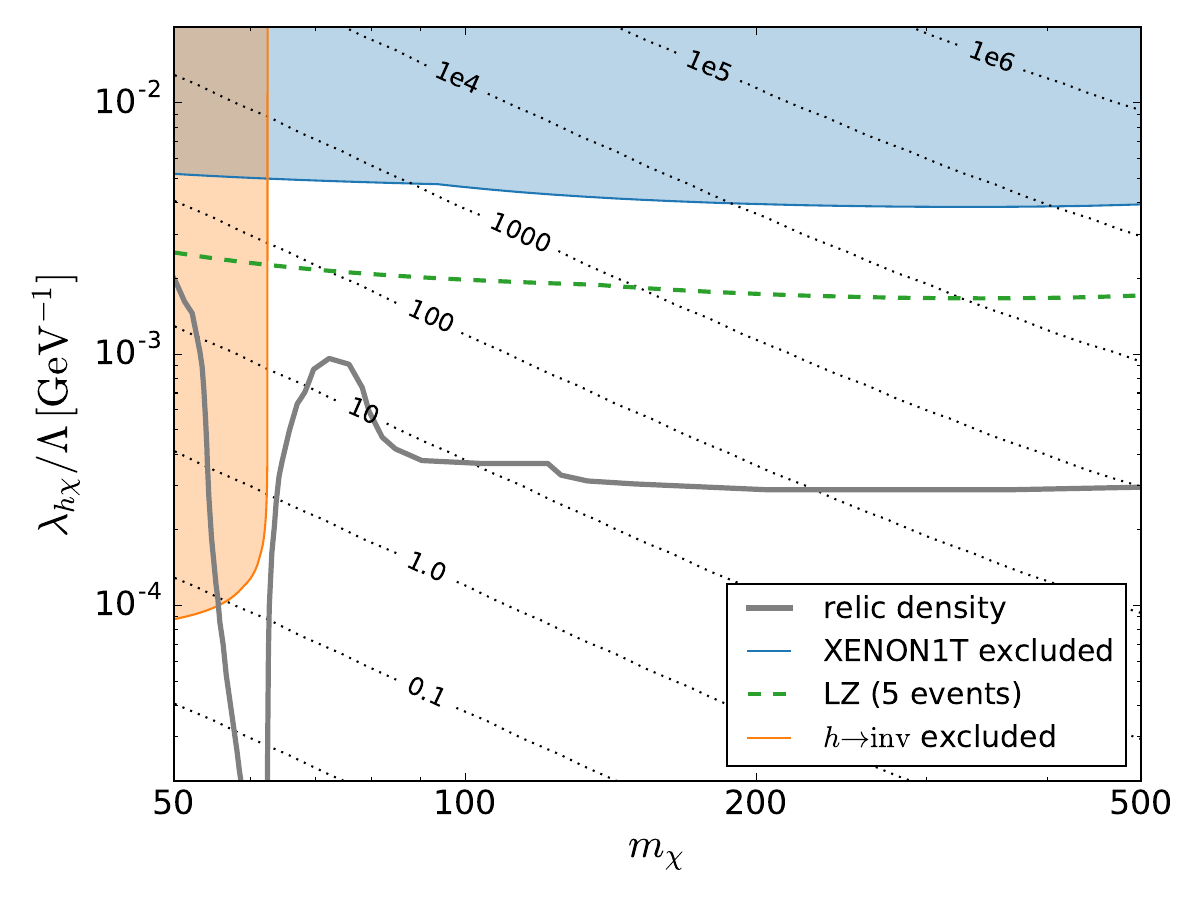}
  \caption{
  Same as figure~\ref{fig:HP} but including the UV-divergent diagrams shown in figure~\ref{fig:UVDivHiggs} for $\Lambda = 1 \, \mathrm{TeV}$. 
  }
  \label{fig:HP_UV}
\end{figure}

In this appendix we provide further details on the additional loop diagrams occurring in the CP-violating Higgs portal model, which are shown in figure~\ref{fig:UVDivHiggs}. Like the diagrams considered in the main text, these diagrams contribute at the order $1/\Lambda^2$ and induce spin-independent interactions for $\phi = \pi/2$. The contribution of these diagrams to the triangle coefficients defined in eqs.~(\ref{eq:trianglecoeff1}) and (\ref{eq:trianglecoeff2}) are given by
\begin{align}
\begin{split}
 \mathcal{C}_S^{\text{triangle}} &\rightarrow \hspace{0.2cm}\mathcal{C}_S^{\text{triangle}} + \frac{g_\chi^2}{(4\pi)^2} \frac{2\,m_\chi}{3\,m_h^2} \left[ (1 + \cos(2\phi))\,B_0(m_\chi^2,\, m_h^2,\,m_\chi^2) + B_1(m_\chi^2,\, m_a^2,\,m_\chi^2)\right]\;,\\
 \mathcal{C}_{PS}^{\text{triangle}} &\rightarrow \hspace{0.2cm} \mathcal{C}_{PS}^{\text{triangle}} + \frac{g_\chi^2}{(4\pi)^2} \frac{2\,m_\chi}{3\,m_h^2}\sin(2\phi)\,B_0(m_\chi^2,\, m_h^2,\,m_\chi^2)\;,
 \end{split}\raisetag{2\normalbaselineskip}
\end{align}
where we have used the replacements from eq.~(\ref{eq:Higgs-replace}). The loop integrals $B_0$ and $B_1$ are UV divergent and we replace the divergences by a logarithmic dependence on the new physics scale $\Lambda$ from eq.~(\ref{eq:LagrCPVHiggs}) according to $1 /\epsilon + \ln(\mu^2/m_\chi^2)\rightarrow \ln(\Lambda^2/m_\chi^2)$~\cite{Haisch:2013uaa}. We study the impact of this additional contribution in figure~\ref{fig:HP_UV} for $\Lambda = 1\,\mathrm{TeV}$. We observe that, while the additional diagrams make the loop-contributions more important, the general conclusions drawn from figure~\ref{fig:HP} are not changed.

\providecommand{\href}[2]{#2}\begingroup\raggedright\endgroup

\end{document}